
\documentclass[aps,twocolumn,superscriptaddress]{revtex4-2}

\usepackage[T1]{fontenc}

\usepackage{color}
\usepackage{soul}
\usepackage[normalem]{ulem}
\usepackage{amsmath,bm}
\newcommand{\stkout}[1]{\ifmmode\text{\sout{\ensuremath{#1}}}\else\sout{#1}\fi}
\usepackage{physics}
\usepackage{amssymb}
\usepackage{bbm}
\usepackage{graphicx}
\usepackage{rotating}
\usepackage{hyperref}
\usepackage{cleveref}
\usepackage{tikz}
\usepackage{algorithm}
\usepackage{algorithmic}
\usepackage{soul}
\usepackage{ dsfont }

\usepackage{multirow}
\usepackage{float}

\usepackage{appendix}

\begin{document}

\title{A language-inspired machine learning approach for solving strongly correlated problems with dynamical mean-field theory}

\author{Hovan Lee}
\email{hovan.lee@rhul.ac.uk}
\affiliation{Department of Physics, Royal Holloway, University of London, Egham, TW20 0EX, United Kingdom}

\author{Zelong Zhao}
\affiliation{King's College London, Theory and Simulation of Condensed Matter (TSCM), The Strand, London WC2R 2LS, UK}

\author{George~H.~Booth}
\affiliation{King's College London, Theory and Simulation of Condensed Matter (TSCM), The Strand, London WC2R 2LS, UK}

\author{Weifeng Ge}
\email{wfge@fudan.edu.cn}
\affiliation{Nebula AI Group, School of Computer Science, Fudan University}

\author{Cedric Weber}
% \email{cedric.weber@kcl.ac.uk}
\affiliation{King's College London, Theory and Simulation of Condensed Matter (TSCM), The Strand, London WC2R 2LS, UK}

\begin{abstract}
We present SCALINN --- Strongly Correlated Approach with Language Inspired Neural Network --- as a method for solving the Anderson impurity model and reducing the computational cost of dynamical mean-field theory calculations. Inspired by the success of generative Transformer networks in natural language processing, SCALINN utilizes an in-house modified Transformer network in order to learn correlated Matsubara Green's functions, which act as solutions to the impurity model. This is achieved by providing the network with low-cost Matsubara Green's functions, thereby overcoming the computational cost of high accuracy solutions. Across different temperatures and interaction strengths, the performance of SCALINN is demonstrated in both physical observables (spectral function, Matsubara Green's functions, quasi-particle weight), and the mean squared error cost values of the neural network, showcasing the network's ability to accelerate Green's function based calculations of correlated materials. 
\end{abstract}

\maketitle

\section{Short Summary}
{We introduce SCALINN, a novel approach that leverages the Transformer architecture—originally designed for natural language processing—to predict Green's functions in strongly correlated systems. The key innovation lies in treating Green's functions as sequential data, akin to language, where the Transformer's ability to capture long-range dependencies in sequences is crucial for modeling discrete frequency dependent Green's functions. Unlike traditional methods, SCALINN does not require explicit knowledge of the underlying physics; instead, it learns to map Green's functions from less demanding computations (e.g., from Hubbard-I or IPT solvers) to accurate, fully correlated outputs. This allows the model to interpolate between different regimes of Green's functions, effectively bridging the gap between low-cost approximations and high-accuracy solutions, which is particularly valuable for solving complex many-body problems in materials science.}

\section{Introduction}

Widespread interest has been devoted in the last three decades to strongly correlated materials, which are being used in emerging technologies, such as spintronics, quantum computing and high temperature superconductivity. They are characterized by strong electronic interactions between their $d$ or $f$-band valence electrons, where the interplay between electron itinerancy (quasi-particle behavior) and electron-electron Coulomb interactions (Mott physics) provides a challenge for standard electronic theories, such as density function theories. The energy scales for the two phenomena overlap for strongly correlated systems. As such, the full characterization of these materials cannot be simplified as the perturbative expansion of one phenomenon against the static backdrop of the another; The competing effects of itineracy and interaction-induced localization necessitates treating these properties on equal footing. 

Through the use of dynamical mean-field theory (DMFT) \cite{dmft_rev_mod}, significant developments have been made towards the understanding of the underlying physics of strong electron correlations in recent years. In particular, the harnessing of DMFT to colloquial materials modeling methods --- such as density functional theory (DFT) with the DFT+DMFT combined approach \cite{dft_dmft} --- has yielded progress into high temperature superconductivity~\cite{jiang2019,PhysRevLett.125.017001,PhysRevB.99.184510}, cold-atom optical trapping~\cite{koepsell2019imaging,brown2019bad,yang2020observation}, and topologically ordered phases~\cite{PhysRevB.103.035125,PhysRevLett.128.043402,PhysRevB.99.045139} among other research topics. In the investigations of these materials, the DFT+DMFT approach provides structural and energetic calculations. These calculations are essential for materials research, particularly when sampling the large phase spaces of material structures with structural relaxation.

Despite these formidable achievements, practical challenges remain with the DFT+DMFT approach. At the heart of the theory is the choice of the quantum engine that solves the many-body Anderson Impurity Model (AIM). This AIM serves as an underlying model that provides the local Green's function --- the descriptor of creation, propagation, and subsequent annihilation of an electron or hole --- of the material of interest via a self-consistent mapping. %In this scheme, the object to be converged is the Green's function: the descriptor of creation, propagation, and subsequent annihilation of an electron or hole in a many-body interacting system. 
Various methods have been developed to calculate the Green's functions of the AIM within the self-consistent cycle of a DMFT calculation. However, each of these solvers have their own limitations. Examples of these solvers include: 

The Hubbard-I (HI) solver~\cite{Hubbard1963ElectronCI}, which assumes no electron itinerancy, and is therefore an approximation that is only reasonable for highly localized systems. 

The iterative perturbation theory (IPT) solver~\cite{RevModPhys.68.13} and its third order extension, which contains all first, second (and optionally third) order irreducible diagrams in the proper self energy. This is an accurate solver in the low correlation regime, but does not generalize well beyond systems that are half-filled and relatively low interaction strength. 

Continuous time quantum Monte Carlo (CTQMC) methods~\cite{gull2007performance,werner2006hybridization,werner2006continuous,rubtsov2005continuous} splits the Hamiltonian of the system into two parts (generally hybridization or interaction), and expands the full partition function as powers of one such part, where these powers are stochastically sampled. While formally exact, this method is nonetheless burdened with the sign problem, random errors, and the requirement to analytically continue the resulting Green's function. 

Finally, the exact diagonalization (ED) solver~\cite{caffarel1994exact,capone2007solving,liebsch2011temperature} which computes the eigenvalues of the AIM Hamiltonian directly, and without approximations in the physics. However, this approach becomes exponentially prohibitive as an increasing number of impurity or bath electron orbitals of the AIM are taken into consideration. This therefore limits the number of orbitals in the calculation, and introduces a finite bath discretization error of the hybridization into the resulting Green's function. 

While this represents an incomplete list of all DMFT solvers that have been considered to date, it is clear that deficiencies remain, and research into improved AIM solvers is still a highly active area of research.

A potential solution could emerge from another discipline; machine learning. Due to advancements in hardware accelerators, improvements in memory capacity, and access to increasingly large databases, deep learning has dominated the field of machine learning in the last decade~\cite{kingma2014adam,lecun2015deep,vaswani2017attention}. These developments have given hope to materials scientists; methods to overcome long standing bottlenecks in condensed matter may be in reach~\cite{dunjko2018machine,carleo2019machine}. A far-from-complete list of these machine learning driven efforts include: predictions of protein structures~\cite{jumper2021highly}, the learning of exchange-correlation functionals for density functional theory~\cite{kirkpatrick2021pushing}, the challenge of ill-conditioned analytic continuation of dynamical quantities~\cite{fournier2020artificial}, or previous efforts aimed at tackling the same problem of DMFT solvers~\cite{arsenault2014machine,sheridan2021data,rogers2021bypassing}. 

In this work, we develop and demonstrate the capabilities of SCALINN --- Strongly Correlated Approach with Language Inspired Neural Network --- which is based on the Transformer architecture~\cite{vaswani2017attention}. SCALINN predicts the Green's functions of strongly correlated systems as ordered sequences in the Matsubara domain, this is done in order to reliably solve single-impurity Anderson models (SIAM) within self-consistent DMFT calculations. Here, prediction refers to the machine learning terminology of the output of a neural network that has been trained on a dataset. 

Rather than encoding the SIAM Hamiltonian directly into the network as input parameters, the characteristics of the SIAM are instead encoded as (potentially multiple) computationally cheap input Green's functions. Optionally, additional characteristic system parameters are also supplemented to the network input, in order to generate more accurate estimates of the target Green's function from the decoder of the transformer. 

The Transformer framework was chosen due to: 1) its potentially infinite memory span --- Green's functions as continuous sequential frequency data exhibit non-local dependence between different frequency points, and therefore the procedural generation of these points require the memory provided by the Transformer network, 2) the parallelism enabled by the Transformer formalism, 3) in the absence of word embedding, the query, key and value method provides a rich representation between Green's function frequency sequence entries. 4) the independence of the model on the number of Matsubara points in training, here the number of points is not case specific, in comparison to previously attempted methods such as fully connected deep neural networks. In contrast, our approach provides predictions of varying lengths without the need of retraining the model, allowing for the learning of abstract information from the dataset of, and the prediction on, materials with a wide range of characteristic energy scales.

Due to these advantages, the Transformer network is able to learn the mapping between computationally cheap, low-level SIAM input Green's functions, and the fully correlated output Green's functions. This enables reliable predictions within DMFT iterations in a non-perturbative manner. Moreover, once the machine learning model has been trained to a satisfactory accuracy (to be defined in section II.B.), the run-time computational cost is independent of the complexity of the problem.

In comparison to other neural networks, a fully connected network or a kernel ridge regression model will only be able to predict series of fixed length. Our model, in contrast, can provide predictions of varying lengths without the need of retraining the model, allowing for the learning of abstract information from the dataset of, and the prediction on, materials with a wide range of characteristic energy scales.

%\textcolor{red}{We showcase the ability of SCALINN}

The performance of SCALINN at different values of temperature and interaction strengths is showcased in this work, along with predictions of spectral functions, quasi-particle weights and Matsubara Green's functions. More technical parameters, including training and testing errors of the network across different hyperparameters are given in the appendix. We report improvements over the team's previous machine learning DMFT solver \cite{d3mft}. In particular, SCALINN is able to predict the Green's function of the Hubbard model at inverse temperature $\beta=100$ (the energy range at which low energy excitation occurs) across various values of Coulomb interaction $U$ with a mean squared error (MSE) of $\sim10^{-7}$, enabling the prediction of the self energy and spectral function of the system. This is a significant improvement from the previous attempt which was able to reach a validation loss of $\sim 10^{-3}$ at $\beta=50$, which was shown to be able to calculate the quasi-particle weight of the system, but was unable to provide predictions on the self energy and spectral function. 

\section{Theory}

\subsection{Dynamical Mean Field Theory}

The SIAM is defined as:
\begin{equation}
\begin{split}
    \hat H_{\mathrm{SIAM}}&=\underbrace{\sum_{i,\sigma}(\varepsilon_{i\sigma}-\mu)\hat a^\dagger_{i\sigma}\hat a_{i\sigma}}_{\hat H_{\mathrm{bath}}}+\underbrace{\sum_{i,\sigma}(V_{i\sigma}\hat c^\dagger_\sigma\hat a_{i\sigma}+h.c.)}_{\hat H_{\mathrm{hyb}}}\\
    &+\underbrace{\sum_{\sigma}(\varepsilon_d-\mu)\hat c^\dagger_\sigma\hat c_\sigma+U\hat c^\dagger_{\uparrow}\hat c_\uparrow\hat c^\dagger_\downarrow\hat c_\downarrow}_{\hat H_{\mathrm{imp}}}\,,
\end{split}
\label{eq:SIAM}
\end{equation}

Where the first term $\hat H_{\mathrm{bath}}$ describes the dispersion of the uncorrelated bath electrons with spin state $\sigma$ and site index $i$, their corresponding annihilation (creation) operators are $\hat a^{(\dagger)}_{i,\sigma}$ and the chemical potential is denoted by $\mu$. The second term $\hat H_{\mathrm{hyb}}$ determines the hybridization between the impurity and the bath electrons, parameterized by the hybridization strength $V_{i\sigma}$. The last term $\hat H_{\mathrm{imp}}$ characterizes the impurity electrons at the single impurity energy level $\varepsilon_d$, with creation (annihilation) operators $\hat c^{(\dagger)}_{\sigma}$ and on-site screened Coulomb interaction strength $U$.

In the absence of interactions, this Hamiltonian can be written in the following block matrix form:

\begin{equation}
    \hat{\underline{\underline H}}^0_{\mathrm{SIAM}}=\begin{bmatrix}
        \varepsilon_d-\mu & \underline V\\
        \underline V^\dagger & \underline{\underline \varepsilon} - \mu \underline{\underline{\mathds{I}}}
    \end{bmatrix}
    \label{eq:SIAMMatrix}
\end{equation}

Due to the absence of interactions, there are no terms that explicitly requires the consideration of different spins, and as such the spin index $\sigma$ is dropped. $\mathbf{\underline V}$ is a vector of the hybridization strengths $V_i$, and $\bm{\underline{\underline \varepsilon}}$ is a diagonal matrix containing the dispersion energies of the uncorrelated bath electrons $\varepsilon_{i}$.

The corresponding non-interacting, total (encapsulating both the impurity and bath parts of the system) Green's function of the system is defined as:

\begin{equation}
\begin{split}
    &(i\omega_n\underline{\underline{\mathds{I}}}-\hat{\underline{\underline H}}^0_{\mathrm{SIAM}})\underline{\underline G}^0_{\mathrm{tot}}(i\omega_n)=\underline{\underline{\mathds{I}}}\,,\\
    &\underline{\underline G}^0_{\mathrm{tot}}(i\omega_n)=\begin{bmatrix}
        i\omega_n+\mu-\varepsilon_d & -\underline V\\
        -\underline V^\dagger & i\omega_n+\mu-\underline{\underline{\varepsilon}}
    \end{bmatrix}^{-1}
\end{split}
\label{Gtot}
\end{equation}

where $i\omega_n$ are the discrete Matsubara frequencies. 

As a side note, although the DMFT procedure can be carried out in real frequency $\omega$, we choose to operate in Matsubara frequency as this approach does not require artificial broadening and does not introduce sharp features in the Green's functions that destabilize convergence. This is in line with methods that interface DMFT with DFT such as TRIQS \cite{parcollet2015triqs}, ABINIT \cite{abinit} and CASTEP \cite{castep}. Most of these implementations are performed in the CTQMC Matsubara frequency framework, as such these solvers scale well with the number of orbitals. On this note, numerical renormalization group (NRG) provides a unique toolset to analyze the spectral features of the single band Hubbard model with a remarkable accuracy. As such, NRG has suggested as a solver for DMFT through various machine learning approaches\cite{andrewmitchell,kim,sturm2021predicting}. However, it is a notorious challenge to extend NRG to multi-orbital systems, and a port of call for DFT+DMFT charge self-consistent approaches lies instead with Matsubara based solvers such as quantum Monte Carlo. %\textcolor{blue}{Note that other projects are ongoing to extend the DFT+DMFT to NRG [cite NRG papers] but those involve some technical challenges [elaborate on challenges] for real material calculations.}

To concentrate on the strongly correlated portion of the system, the characterization of the impurity electrons does not require a complete description of the bath electron dispersions or the hybridization strengths. As such, the bath information can be condensed; taking only the top left block of $G^0_{\mathrm{tot}}(i\omega_n)$, which is denoted as the non-interacting impurity Green's function $G^0_{\mathrm{imp}}(i\omega_n)$, this $G^0_{\mathrm{imp}}(i\omega_n)$ can be analytically calculated via block-wise inversion of \cref{Gtot}:

\begin{equation}
    G^0_{\mathrm{imp}}(i\omega_n)=\frac{1}{i\omega_n+\mu-\varepsilon_d-\Delta(i\omega_n)}\,,
    \label{eq:NonIntImpGF}
\end{equation}
\begin{equation}
    %\textcolor{red}{\Delta(i\omega_n)=V_i\left(\frac{1}{i\omega_n+\mu-\varepsilon_{ij}}\right)V^*_j\,,}
    \Delta(i\omega_n)=\sum_i\frac{|V_i|^2}{i\omega_n-\varepsilon_i}
    \label{eq:delta}
\end{equation}

% \textcolor{red}{where $G^0_{\mathrm{imp}}(i\omega_n)$ is the non-interacting impurity Green's function, and is the top left block of $G^0_{\mathrm{tot}}(i\omega_n)$. $\Delta(i\omega_n)$ is the hybridization function, which encapsulates all contributions of the bath electrons to the properties of the impurity. In this project, this hybridization function is approximated with finite bath discretization:}

where $\Delta(i\omega_n)$ is known as the hybridization function or the dynamical mean field.

In ED, this $\Delta(i\omega_n)$ is approximated with a finite number of bath orbitals:

\begin{equation}
    \Delta(i\omega_n) \approx \Delta^{\mathrm{bath}}(i\omega_n)=\sum^{N_b}_{p=1}\frac{V^2_p}{i\omega_n-\varepsilon_p}\,,
    \label{eq:delta2}
\end{equation}

with number of bath sites $N_b$ and bath site index $p$. Each bath site has hybridization strength $V_p$ and energy $\varepsilon_p$. These parameters are typically fit via numerical techniques \cite{mejuto2020efficient}.

The effects of interactions is then supplemented to $G^0_{\mathrm{imp}}(i\omega_n)$ via the Dyson equation to obtain the (interacting) impurity Green's function $G_{\mathrm{imp}}(i\omega_n)$:
\begin{equation}
    [G_{\mathrm{imp}}(i\omega_n)]^{-1}=[G^0_{\mathrm{imp}}(i\omega_n)]^{-1}-\Sigma(i\omega_n)\,,
    \label{eq:Dyson}
\end{equation}
where $\Sigma(i\omega_n)$ is the self-energy term given by the DMFT solver.

The resulting $G_{\mathrm{imp}}(i\omega_n)$ is approximated as the wavevectors $\underline k$ averaged Green's function of the Hubbard model:
\begin{equation}
    \hat H_{\mathrm{Hubbard}}=-t\sum_{<i,j>,\sigma}c^\dagger_{j,\sigma}c_{i,\sigma}+U\sum_ic^\dagger_\uparrow c_\uparrow c^\dagger_\downarrow c_\downarrow - \mu\sum_{i,\sigma}c^\dagger_\sigma c_\sigma\,,
    \label{eq:Hubbard}
\end{equation}

where $c^{(\dagger)}_{i,\sigma}$ denotes the annihilation (creation) operator on the site index $i$ of spin state $\sigma$, $t$ is the hopping energy between neighboring sites, $U$ is the on-site screened Coulomb interaction strength, and $\mu$ is the chemical potential.

With periodic boundary conditions, the Green's function of the Hubbard model is can be presented in terms of wavevectors $\underline k$, which approximates to the SIAM interacting impurity Green's function via:
\begin{equation}
    G_{\mathrm{imp}}(i\omega_n)\approx \frac{1}{N_{\underline k}}\sum_{\underline k}^{N_{\underline k}}G^{\mathrm{Hubbard}}_{\underline k}(i\omega_n)\,,
    \label{eq:Approx}
\end{equation}

where $G^{\mathrm{Hubbard}}_{\underline k}(i\omega_n)$ is the $\underline k$ dependent Hubbard model Green's function, and $N_{\underline k}$ is the number of $\underline k$ points taken into consideration.

In short, \cref{eq:SIAM,eq:SIAMMatrix,Gtot,eq:NonIntImpGF,eq:delta,eq:delta2,eq:Dyson,eq:Hubbard,eq:Approx} provides a pathway to approximating the Green's function of the Hubbard model --- a model that is colloquially used to describe the transition between the insulating and conducting phenomena of strongly correlated materials --- with the simpler SIAM, providing that the parameters of the SIAM are well chosen.

\begin{figure}
    \centering
    \includegraphics[width=\linewidth]{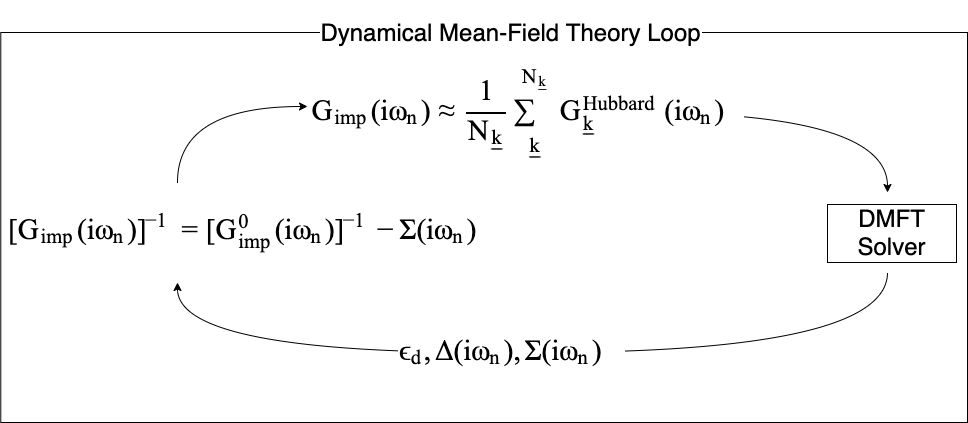}
    \hrule
    \caption{\textbf{Schematic diagram of Dynamical mean-field theory (DMFT) iterative loop.} To initiate the DMFT calculation, a set of initial parameters are chosen for the impurity energy level $\epsilon_d$, the hybridization function $\Delta(i\omega_n)$, and the self-energy $\Sigma(i\omega_n)$ (as denoted at the bottom of the figure). The non-interacting impurity Green's function $G_{\mathrm{imp}}^0(i\omega_n)$ is evaluated via \cref{eq:NonIntImpGF}, which is then used to calculate the impurity Green's function $G_{\mathrm{imp}}(i\omega_n)$ with \cref{eq:Dyson} (left of figure). Next, the local Green's function of the Hubbard model $\frac{1}{N_{\underline k}}\sum_{\underline k}^{N_{\underline k}}G^{\mathrm{Hubbard}}_{\underline k}(i\omega_n)$ is approximated with $G_{\mathrm{imp}}(i\omega_n)$ (top of figure). The DMFT solver is the applied to provide a new set of $\epsilon_d, \Delta(i\omega_n)$, and $\Sigma(i\omega_n)$ parameters to be used in the next iteration of the DMFT loop. This process is carried out until convergence.}
    \label{fig:DmftLoop}
\end{figure}

In DMFT, the choice of the parameters of the SIAM are facilitated through the use of the DMFT solver, and the DMFT self-consistency cycle: a set of initial SIAM parameters are chosen, calculations of $G_{\mathrm{imp}}(i\omega_n)$ are performed, this Green's function is then approximated to the Hubbard model Green's function, which is then fed into the DMFT solver to provide a new updated set of parameters for the SIAM, as shown in \cref{fig:DmftLoop}.

Updating this self-consistent loop until convergence is a well defined but demanding task, and is where the overhead of computational cost sits. Conventionally this task is carried out by the DMFT solver, here SCALINN is applied to approximate the iterative update of $G_{\mathrm{imp}}(i\omega_n)$ with near instantaneous overhead.
%{This is where the solver comes in ... expand on this (this is a very well defined but demanding task, where the overhead sits) here scalinn is going to be another approximation, however with near instantaneous overhead. Plus update figure}

\subsection{The SCALINN Solver}

\subsubsection{Modes of Operation}\label{subsec:modes}

\begin{figure}
    \centering
    \includegraphics[width=\linewidth]{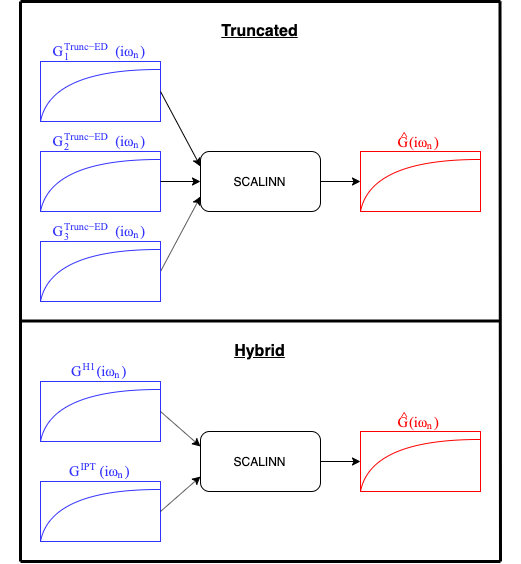}
    \caption{\textbf{Illustration of the modes of operation in SCALINN. }Input (output) Green's functions are depicted in blue (red). Top: the Truncated mode --- A set of three Green's functions calculated from the exact-diagonalization solver with a number of bath sites discarded are used as inputs. Bottom: the Hybrid mode --- Input correspond to Green's functions calculated from the Hubbard I and the iterative perturbation theory solvers. Both modes of operation result in a prediction Green's function $\hat G(i\omega_n)$.}
    \label{fig:modes}
\end{figure}

We considered two different modes of operation for the SCALINN solver as shown in \cref{fig:modes}:

The first corresponds to the potential to overcome bath discretization error in ED solvers; due to the growth in the dimensions of the Hamiltonian as the number of bath sites increases, we attempted to train networks that would reconstruct the Green's function of $N_b$ bath site calculations by providing the network with a set of ED solver calculated Green's functions in which a number of the $N_b$ bath sites are discarded. We call this the `truncated' mode of operation.

Whilst we have trained models from various different numbers of bath sites. Due to the computational cost to generate the training datasets, we report our results from a model trained on 7-bath sites in the main text, with details and results of the other models in the appendix. We report that as the number of bath sites used to train the model increases, the accuracy of predictions increases.

Given the exponential scaling in calculations with respect to bath size, this method would provide substantial benefit in computational cost. Three of these truncated systems were created, such that all of the 7 bath orbitals from the full system is present in at least one of the truncated systems. This is a similar motivation to the `distributed exact diagonalization' approach \cite{granath2012distributional}. The mapping between the Green's functions of these truncated systems and the full 7 bath site Green's function is then learned by the network. We report that is approach produces Green's functions that are in agreement with the exact calculations obtained from CTQMC.

The second approach is for the network to learn $G^{ED}$ from input Green's functions of complementary characteristics. In our attempts, the HI and the IPT solvers were used. These two input models, while computationally cheap, allow for a description of different extremes in correlated physics. Specifically, HI is superior in the atomic limit, while IPT describes the low-$U$ itinerant physics as a perturbative expansion in powers of $U$. We will denote this approach the `hybrid' method.

\subsubsection{Encoder-Decoder Structure \& Auto-regression}

\begin{figure}
    \centering
    \includegraphics[width=\linewidth]{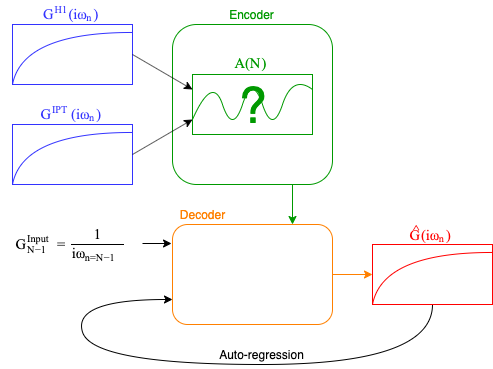}
    \caption{\textbf{Illustration of the transformer framework in SCALINN: }Input Green's functions (blue) are mapped to some abstract sequence $A(N)$ by the encoder (green). The decoder (orange) takes the input $G^{\mathrm{input}}_{N-1}=\frac{1}{i\omega_{n=N-1}}$, together with the encoder output, to auto-regressively generate prediction (red) tokens $\hat G$ that form the prediction Green's function $\hat G(i\omega_n)$.}
    \label{fig:encoder-decoder}
\end{figure}

As colloquially used in natural language processing, the transformer architecture \cite{vaswani2017attention} constitutes an encoder-decoder structure to generate output sequences from input prompts; The encoder maps the input prompt sequences to abstract sets of numerical sequences, whereas the decoder takes and converts the abstract sequences into output tokens. This is illustrated in \cref{fig:encoder-decoder}

In the case of the SCALINN solver with the `hybrid' approach, the encoder takes input Green's function sequences of $N$ decreasing Matsubara frequencies: the Green's function calculated with the Hubbard-I solver $G^{\mathrm{H1}}(i\omega_n)=\left(G^{\mathrm{H1}}_{N-1}, G^{\mathrm{H1}}_{N-2}, \dots, G^{\mathrm{H1}}_{0}\right)$ --- where $G^{\mathrm{H1}}_{N-1}$ represents the Green's function value at the highest Matsubara frequency considered, $G^{\mathrm{H1}}_{N-2}$ is the value at the next highest Matsubara frequency, etc. --- and $G^{\mathrm{IPT}}(i\omega_n)$, a Green's function calculated from the iterative perturbation theory (IPT) solver with the same decreasing frequency order as $G^{\mathrm{H1}}(i\omega_n)$. The network then converts these sequences into abstract sequences $A(N) = (a_{N-1}, a_{N-2}, \dots, a_{0})$.

Similarly for the `truncated' approach, instead of $G^{\mathrm{H1}}(i\omega_n)$ and $G^{\mathrm{IPT}}(i\omega_n)$ input Green's functions, we supply the encoder with Green's functions calculated from the truncated systems. Apart from this difference the methodology remains the same, in order to avoid the explanations of the same concepts twice, the remainder of this section will continue with the explanation of the `hybrid' mode of operation.

Next, an input corresponding to the guess of the Green's function value at the highest frequency considered $G^{\mathrm{input}}_{N-1}=\frac{1}{i\omega_{n=N-1}}$ is supplied to the decoder. This input, along with the abstract sequences $A(N)$ (which are themselves mapped onto a set of attention vectors as described below) are then mapped by the decoder into the predicted token corresponding to the Green's function value at the next Matsubara frequency $\hat G_{N-2}$.
% In the case of the SCALINN solver, the encoder takes an input Green's function sequence of $n'$ decreasing Matsubara frequencies $G^{\mathrm{input}}(i\omega_n)=\left(G_N, G_{N-1}, \dots, G_{N-(n'-1)}\right)$ --- where $G_N$ represents the Green's function value at the highest Matsubara frequency considered, $G_{N-1}$ is the value at the next highest Matsubara frequency, etc. --- and converts it into an abstract sequence $A(n) = (a_{N}, a_{N-1}, \dots, a_{N-(n'-1)})$. This abstract sequence is then mapped by the decoder into the predicted token corresponding to the Green's function value at the next Matsubara frequency $\hat G_{N-n'}$.

The decoder then generates the next predicted token $\hat G_{N-3}$ by taking $A(N)$ (in their attention vector form), $G^{\mathrm{input}}_{N-1}$, and $\hat G_{N-2}$ as inputs. The decoder carries out this process auto-regressively, generating $\hat G$ tokens from $G^{\mathrm{input}}_{N-1}$, $A(N)$ and all previous output tokens, until the prediction at the lowest frequency $\hat G_0$ is reached.

% \textcolor{red}{Physically, the use of auto-regression allows us to build a prediction method that makes use of the }

{The significance of the use of auto-regression with descending frequency is that the network begins its prediction at the limit of high-frequency. At this limit, the Green's function describes single particle excitations. When expanded in this high frequency limit, the leading term of physical Green's functions scales as $1/i\omega_n$, which is the value we make use of as the input token $G^{\mathrm{input}}_{N-1}$.}

{As the network iterates towards predictions at lower and lower frequencies, the referencing of the $G^{\mathrm{H1}}(i\omega_n)$ and $G^{\mathrm{IPT}}(i\omega_n)$ sequences, as well as the training of the network, becomes more and more consequential. This provides a systematic approach from the single election ionization physics, towards the intermediate and the low frequency regimes that describe phenomena such as the Mott transition and Fermi liquid behavior.}

\subsubsection{Self-Attention Mechanism}

In order to map and manipulate the abstract sequences, the transformer architecture makes use of the self-attention mechanism --- a method which splits each token in a sequence (in this case these tokens would be taken from the abstractly coded sequences of $G^{\mathrm{H1}}(i\omega_n), G^{\mathrm{IPT}}(i\omega_n), G^{\mathrm{input}}_{N-1}$ and the sequence of generated $\hat G$ tokens) into three learned vectors: the query, key and value vectors.

Whilst the naming of the query, key and value vectors is not immediately important in this work, a conceptual understanding of these vectors can be grasped in the context of information retrieval and database query processing. In these domains, queries represents user input statements of some desired information. However, a query does not uniquely identify a single result in the database, and is instead referenced with the key descriptors of the objects in the database in order to return a ranked list of results based on the similarities between the query and the keys. The value of the objects correspond to the data content of each object.

{For physical Green's functions, we know that successive points in $G(i\omega_n)$ are complementary to one another --- $G(i\omega_n)$ are smooth and continuous, reflecting the fact that the system's excitations (e.g., electrons, holes, or quasiparticles) are not isolated events but are part of a continuous spectrum of states. Moreover, emergent effects (e.g., Mott transitions or Kondo physics) introduce correlations that spans all frequencies scales. For example, the formation of a Mott gap at low frequencies is influenced by the high-frequency structure of the self-energy. We make use of the attention mechanism to handle these dependencies between Matsubara frequency points.}

In the case of SCALINN, a set of query, key and value vectors are evaluated for each token based on the training of the network. The length $d_k$ query vector of a specific token is to be multiplied with the length $d_k$ key vectors of all the tokens in the sequence via dot product. This is done in order to calculate the similarities of the query vector of the specific token with the key vectors of the tokens in the sequence. The output of the self-attention mechanism is then a weighted sum where the value vectors of the token are weighted by the dot product between the query and key vectors.

When vectorized across a length $n'$ input sequence, this dot product would be expressed as $\underline{\underline Q}.\underline{\underline K}^T$, where $\underline{\underline Q}$ is a $n'\times d_k$ matrix in which the $n'$ rows correspond to the length $d_k$ query vectors of the tokens, $\underline{\underline K}^T$ is a $d_k\times n'$ matrix where the $n'$ columns correspond to the length $d_k$ key vectors of the tokens. This $n'\times n'$ matrix $\underline{\underline Q}.\underline{\underline K}^T$ is then scaled by $1/\sqrt{d_k}$, and subsequently applied to a softmax function. This softmax function normalizes the rows of $\underline{\underline Q}.\underline{\underline K}^T/\sqrt{d_k}$, such that the $n_i$-th row correspond to a normalized set of $n'$ metrics based on the similarities between the query vector of the $n_i$-th token and the key vectors of the tokens across the whole sequence: This set of metrics are the attention weights, corresponding to how much attention should be placed on the each of the $n'$ tokens given the query of the $n_i$-th token.

A dot product is then applied between this attention weight matrix and the $\underline{\underline V}$ matrix of size $n'\times d_v$, such that the attention values are applied as the weights of a weighted sum to the value vectors of the sequence. 

The full operation is expressed as:
\begin{equation}
    \mathrm{Attention}(Q,K,V)=\mathrm{softmax}\left(\frac{\underline{\underline M}+\underline{\underline Q}.\underline{\underline K}^T}{\sqrt{d_k}}\right).\underline{\underline V}\,,
    \label{eq:masked_attention}
\end{equation}
where $\underline{\underline M}$ is a $n'\times n'$ matrix of mask values, and is described below.

\begin{figure*}
    \centering
    \includegraphics[width=0.8\linewidth]{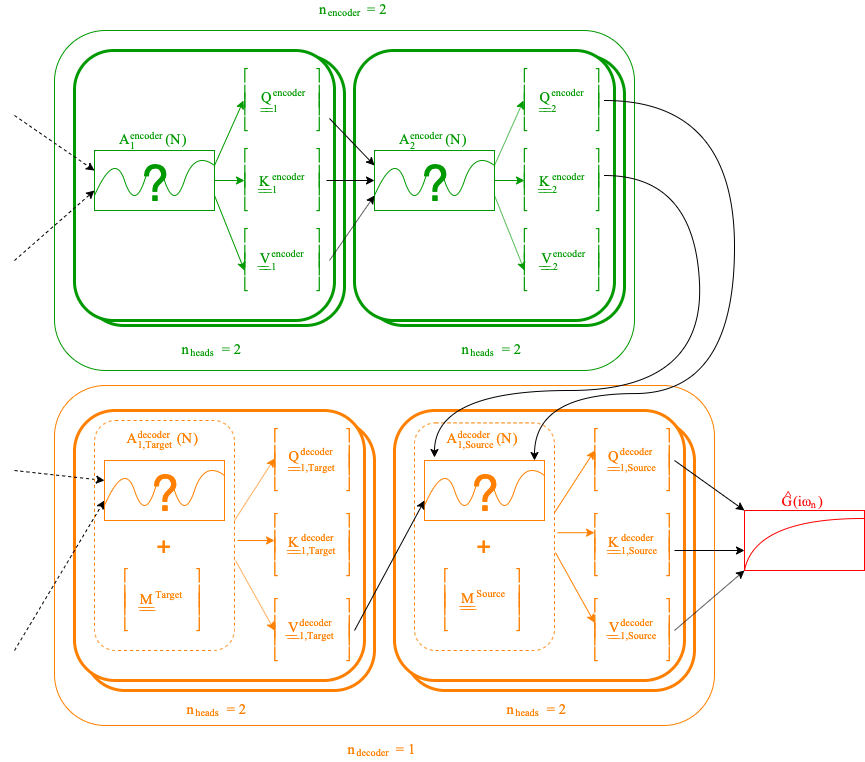}
    \caption{\textbf{Illustration of masking and the self-attention mechanism: }dotted lines denote the inputs to the encoder (green) and the decoder (orange). The inputs of the encoder are mapped into the abstract sequence $A_1^{\mathrm{encoder}}(N)$. In this example, the hyperparameter $n_{\mathrm{encoder}}=2$ was chosen, giving rise to two successive sets of self-attention query, key and value vectors in the encoder, denoted as $\underline{\underline Q}^{\mathrm{encoder}}_i,\underline{\underline K}^{\mathrm{encoder}}_i,\underline{\underline V}^{\mathrm{encoder}}_i$ respectively, for $i=1,2$. The query, key and value vectors of the first encoder block is then mapped to the abstract sequence for the next encoder block $A_2^{\mathrm{encoder}}(N)$. From $A_2^{\mathrm{encoder}}(N)$, the second encoder set of query, key and value vectors are obtained. Likewise, the input to the decoder are mapped to $A_1^{\mathrm{decoder}}(N)$, onto which a target mask $\underline{\underline M}^{\mathrm{Target}}$ is applied in order to calculate the decoder target query, key and value vectors. The decoder source block then takes the last set of encoder query and key vectors $\underline{\underline Q}_2^{\mathrm{encoder}},\underline{\underline K}_2^{\mathrm{encoder}}$, along with the target decoder value vectors $\underline{\underline V}_{1,\mathrm{Target}}^{\mathrm{decoder}}$ in order to obtain the abstract decoder source sequence $A^{\mathrm{decoder}}_{1,\mathrm{Source}}(N)$. From $A^{\mathrm{decoder}}_{1,\mathrm{Source}}(N)$ the source decoder query, key and value vectors are calculated in order to predict the Green's function $\hat G(i\omega_n)$. All encoder and decoder sub-blocks are depicted with $n_{\mathrm{heads}}=2$ heads to showcase the parallelism of the method.}
    \label{fig:enter-label}
\end{figure*}

\subsubsection{Masking \& Mastubara Encoding}
Despite handling the input prompts as sequences, the transformer network does not parse inputs and previously predicted tokens sequentially. Instead this process is carried out in parallel: each head of the multi-head attention blocks employed in the network applies its own version of the attention mechanism of \cref{eq:masked_attention}. 

Generally, with a higher number of heads, the richness of information that can be learned in these multi-head self-attention blocks increases. {For example, a number of heads may be dedicated to capture the local (in terms of energy) smooth and continuous nature of Green's functions, whilst other heads may capture many-body effects that spans the whole frequency spectrum.} The hyperparameter corresponding to the number of heads is varied, and the performances of models with different number of heads, and different number of decoder blocks ($n_{Decoder}$) are listed in \cref{tab:beta100_truncated5}.

As such, the order of the input sequences in descending Matsubara frequency must be encoded into the tokens of the sequence explicitly. Our attempts at various different Matsubara encoding schemes are detailed in the appendix.

This non-sequential in manner of sequence-handling however, would mean that during training, the self-attention heads of the network are able to reference full length $N$ target Green's function sequences $G^{\mathrm{Training}}(i\omega_n)$ obtained from the exact diagonalization solver for $7$ bath sites. These $G^{\mathrm{Training}}(i\omega_n)$ correspond to the values that a fully trained network would be able to predict (i.e. the predicted Green's function $\hat G=(\hat G_{N-1},\hat G_{N-2},\dots,\hat G_0)$). Whereas, the network should be trained such that it cannot reference $\hat G$ tokens of lower frequencies that it has not generated yet.

As such, mask matrices $\underline{\underline M}$ are applied to the self-attention blocks in the decoder, in order to mask these lower frequency tokens such that the network will not be able to reference them during training.

The decoder contains two types of self-attention blocks: Target self-attention blocks (corresponding to the blocks which, with $n_{\mathrm{Decoder}}=1$, handles only the self-attention of the prediction --- target, in the jargon of machine learning --- tokens $\hat G$ that are generated), and source self-attention blocks (which takes the query and key vectors from the input --- source --- encoder block and the value vectors from the target self-attention block).

For these two types of self-attention blocks, we trialed different masked matrices (as detailed in the appendix) during training of the following form:
\begin{equation}
\begin{split}
    &M^{\mathrm{Target}}_{ij}=\begin{cases}
        0,&\text{if}\,\,i-LB<j<i+1\\
        -\infty,&\text{otherwise}
    \end{cases}\,,\\
    &\underline{\underline M}^{\mathrm{Target}}_{LB=2}=\begin{bmatrix}
        0&-\infty&-\infty&\dots&\dots\\
        0&0&-\infty&-\infty&\dots\\
        -\infty&0&0&-\infty&\dots\\
        \vdots&\ddots&\ddots&\ddots&\ddots\\
        -\infty&\dots&-\infty&0&0
    \end{bmatrix}\,,
\end{split}
\end{equation}
for the mask of the target self-attention block, where $LB$ is the look-back hyperparameter, which denotes the number of previous tokens minus one in the sequence that the network can reference. This acts as a Matsubara frequency window that determines which entries in the sequence can influence the prediction of the token being generated.

For example with $LB=2$, when predicting the first token $\hat G_{N-2}$ to continue the prompt of $G^{\mathrm{input}}_{N-1}$, the first row of $\underline{\underline M}^{\mathrm{Target}}$ renders the network unable to reference $G^{\mathrm{Training}}(i\omega_n)$ for $n<N$. When predicting the next token $\hat G_{N-2}$, the network is able to reference $\hat G_{N-2}$, and is not able to reference $G^{\mathrm{Training}}(i\omega_n)$ for $n<N-1$. When predicting $\hat G_{N-3}$, it is able to reference $\hat G_{N-2}$, and so forth.

For the source self-attention block, we would like the block to be able to reference the $G^{\mathrm{IPT}}(i\omega_n)$ and $G^{\mathrm{H1}}(i\omega_n)$ around the frequency that the network is predicting, as such the form of the mask matrices applied is:
\begin{equation}
\begin{split}
    &M^{\mathrm{Source}}_{ij}=\begin{cases}
        0,&\mathrm{if}\,\,i-LB<j<i+1+LF\\
        -\infty,&\mathrm{otherwise}
    \end{cases}\,,\\
    &\underline{\underline M}^{\mathrm{Source}}_{LB=2,LF=1}=\begin{bmatrix}
        0 & 0 & -\infty & -\infty & \dots & \dots\\
        0 & 0 & 0 & -\infty & \infty & \dots\\
        -\infty & 0 & 0 & 0 & -\infty & \dots\\
        \vdots & \ddots & \ddots & \ddots & \ddots & \ddots \\
        -\infty & \dots & -\infty & 0 & 0 & 0 \\
        -\infty & \dots & \dots & -\infty & 0 & 0
    \end{bmatrix}\,,
\end{split}
\end{equation}
such that when generating the $\hat G_{n'}$ token, the network is able to reference $\{G^{\mathrm{H1}}_{n''}\}$ and $\{G^{\mathrm{IPT}}_{n''}\}$ tokens, where $n'-(LB-1)\le n'' \le n'+LF$.

The consequence of the application of these masks is as follows: the dots product calculations on the similarities of the query and key vectors in \cref{eq:masked_attention} corresponding to the $n_i$ entries to be masked is numerically reduced to $-\infty$ via the addition $\underline{\underline M}+\underline{\underline Q}.\underline{\underline K}^T$, and as such the softmax function does not consider the masked $n_i$ entries in the normalization of attention weights; the resulting weights at these masked indices are 0.

\subsubsection{Training vs. Run-time}

As alluded to, there is a slight difference of the network during training and during run-time:

In training the network, the decoder is sequentially provided with the 7 bath site ED Green's function values starting at the highest frequency considered. This is done in order for the network to generate a prediction token for the Green's function value at the next highest frequency. 

The cost function to be minimized is defined as the mean squared error to the 7-bath ED solution at a set of training points, 
\begin{equation}
    J(\hat{G}(i\omega_n),G^{\text{ED}}(i\omega_n))=\frac{1}{N}\sum^{N-2}_{n=0}(\hat{G}(i\omega_n)-G^{\text{ED}}(i\omega_n))^2\,,
\end{equation}
where $\hat{G}(i\omega_n)$ is the indexed set of predicted tokens $\hat G$, $G^{\text{ED}}(i\omega_n)$ is the 7-bath ED solution of the SIAM, and the sum is performed over Matsubara frequency points of the generated tokens. $N$ can be increased such that the model is trained to generate as many Matsubara frequency points as required. In practice, increasing the number of Matsubara points considered beyond $N=32$ in this mean squared error evaluation does not dramatically increase the accuracy of the prediction, as will be shown below, the results obtained from the model is within the statistical error bars of the CTQMC result.

During run-time however, only the highest frequency point is provided to the decoder with $G^{\mathrm{input}}_{N-1}=\frac{1}{i\omega_{n=N-1}}$ as mentioned above. To continue past the first prediction, the subsequent generated tokens are provided back to the decoder auto-regressively.

\section{Results}

\subsubsection*{Computational Details}
To setup the training data of the model, three $G^{\text{ED}}(i\omega_n)$ data-sets at different inverse temperatures of $\beta=10,50,100$ were generated, each consisting of 14,000 different 7-bath SIAM Hamiltonians. The ED Green's functions were generated using the ED-KCL package~\cite{weber2012augmented}, with DMFT functionality where required enabled via interface to the TRIQS package ~\cite{parcollet2015triqs}. Implementation of the Transformer model made use of the PyTorch package~\cite{paszke2017automatic}. The training of our base model, detailed in Table \ref{tab:beta100_truncated5}, was conducted on a single NVIDIA A100 GPU, where each model was trained on 80\% of the data for 350,000 steps, which took approximately 9 hours per model, with the rest of the data used for evaluation of the model accuracy. During run-time, each DMFT loop takes $\sim10^3$ seconds on the A100 GPU. It should be noted that this GPU computational cost is independent on the number of bath sites of the training dataset. In contrast, a traditional 7 bath site ED calculation on the same GPU takes $\sim 1$ minute, the size of the Green's function matrix to be inverted (as shown in \cref{fig:DmftLoop}) scales as $2^N$ for a paramagnetic calculation, and as such the computational cost of each ED DMFT loop scales accordingly.

The bath parameters of the SIAM for these data-sets were calculated as approximations to a continuous semi-circular spectral hybridization,
\begin{equation} \label{eq:semi-circular}
\Delta^{\text{Bethe}
} \left(i \omega_n\right)=\frac{1}{2 \pi^2} \int_{-\infty}^{\infty} d \omega \frac{\sqrt{W^2-\omega^2}}{i \omega_n-\omega} \Theta(W-|\omega|)\,,
\end{equation}
with bandwidth $2W$. This hybridization represents the paradigmatic Bethe lattice, which represents an exact model for DMFT~\cite{georges1992hubbard}, with physical hybridizations of relevance to the applicability of DMFT expected to be close to this form. The different 7-bath Hamiltonian parameters were found by initializing the parameters randomly, and subsequently minimizing the error in the effective hybridization of the model (~\cref{eq:delta2}) via changes in $V_p$ and $\epsilon_p$ according to

\begin{equation}
    \sum_{i\omega_n}\frac{1}{i\omega_n}(\Delta^{\text{bath}}(i\omega_n)-\Delta^{\text{Bethe}}(i\omega_n))^2 .
\end{equation}
This fit is constraint to ensure that particle-hole symmetry of the bath parameters is maintained (with one bath energy constrained to $\omega=0$), and is stopped early when the squared error in the fit reaches only $1W$, in order to provide the different Hamiltonians. To complete the definition of the SIAM, $U$ values are uniformly generated in the range 0-10$W$.

The impurity level $\epsilon_d$ is kept at $-U/2W$, and the chemical potential $\mu$ is set to $-\epsilon_d - U/2W$. Where these values were chosen to ensure consideration of the particle-hole symmetric point of these Hamiltonians. When demonstrating the use of SCALINN as a solver in a DMFT calculation, the self-energy of ~\cref{eq:Dyson} is self-consistently updated, with the interactions able to induce significant changes to the approximately semi-circular initial spectrum, including metal-insulator phase transitions.

\subsection{Truncated Hamiltonian}

\begin{figure}[h!]
    \centering
    \includegraphics[width=1.\linewidth]{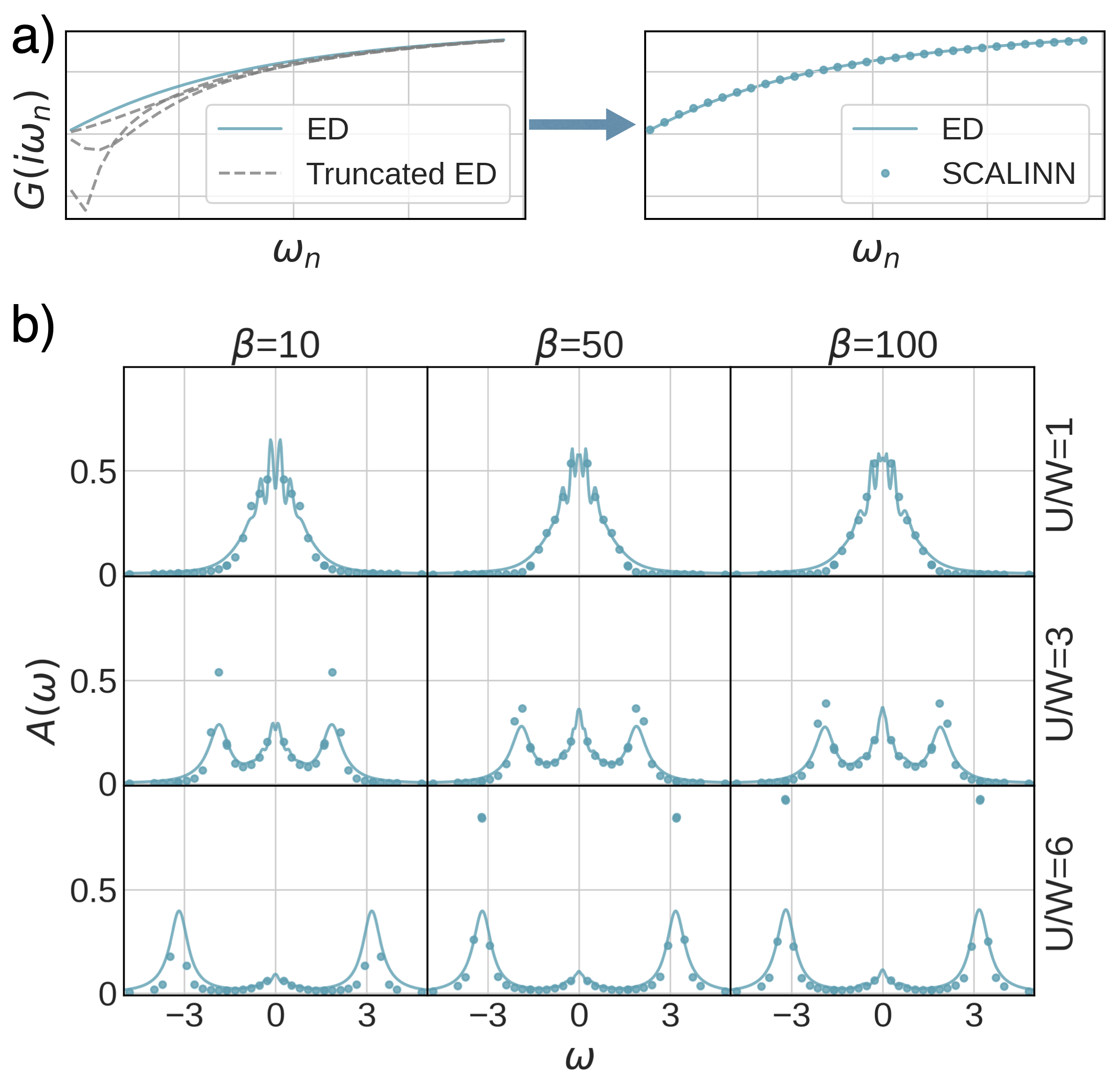}
    \caption{\textbf{Representative SCALINN predictions for a test-set SIAM with the truncated bath scheme.} In a), the desired (7-bath) ED solution $G^{\text{ED}}(i\omega_n)$ is plotted as a blue line, and the three truncated (3-bath) $G^{Trunc}(i\omega_n)$ are plotted in dashed gray. These $G^{Trunc}(i\omega_n)$ act as inputs to the Transformer model to produce the SCALINN predictions $\hat G(i\omega_n)$, which are plotted as blue dots. In b), Pad\'e analytic continuation obtains the predicted spectral function $A(\omega)$, shown here for three different values of $\beta$ and $U$. Once again, the ED solution is plotted as blue lines, and the SCALINN predictions are plotted as blue dots.}
    \label{fig:truncated_aw}
\end{figure}

We first consider the `truncated' mode of operation of the SCALINN model, whereby predictions of the 7-bath SIAM models are created from ED Green's functions with only 3 bath orbitals, mitigating the exponential increase in cost of ED with respect to bath size. The 3-bath SIAM Hamiltonians were created from their 7-bath counterparts by inclusion of the $\epsilon=0W$ bath orbital, and then selection of one particle-hole symmetric pair of bath orbitals at finite frequency. This allows us to create three 3-bath SIAM approximations to each SIAM of interest. These are solved with ED to provide the input to SCALINN, as described in ~\cref{subsec:modes}.

An example of the effect that this bath dropout has on Green's functions is shown in \cref{fig:truncated_aw}.a, where these truncated bath models differ significantly from the desired 7-bath solution. Unless otherwise specified, $G(i\omega_n)$ indicates only the imaginary component of the Green's functions in all the figures shown below.

SCALINN nevertheless predicts the 7-bath ED solution with high accuracy after training of the model. SCALINN models are trained at three different inverse temperatures, with the average training error reduced to below $J(\hat{G}(i\omega_n),G^{\text{ED}}(i\omega_n))<10^{-5}$ in all cases. We can analytically continue the predicted Matsubara Green's functions onto the real-frequency axis via Pad\'e approximants, to consider their accuracy of the real-frequency spectrum, shown in \cref{fig:truncated_aw}.b for a representative test-set prediction, with three values of inverse temperature $\beta$ and three values of interaction $U$.

The SCALINN predictions agree very well with the ED ground truth for these systems, even on the real axis, with only small deviations are observed in \cref{fig:truncated_aw}.b.
As the interaction increases, classic hallmarks of correlated materials emerge, with low $U$ describing metals with a single quasi-particle peak at zero energy. As temperature increases (or as $\beta$ decreases), so does the rate of scattering between electrons, leading to a broadening of these peaks. At intermediate interaction strength $U=3W$, lower and upper Hubbard bands at $\pm U/2W$ emerge around the quasi-particle peaks at $\omega=0$. Once again, due to increasing rates of scattering, these three sets of peaks are broadened as temperature increases. Lastly, at higher interaction strength still $U=6W$, the magnitude of the quasi-particle peaks are greatly reduced in favor of the Hubbard bands. However, due to the lack of DMFT self-consistency in this case, the fully insulating Mott solution was not recovered.

\subsection{Hybrid Approximation Solver}

\begin{figure}[h!]
    \centering
    \includegraphics[width=1.\linewidth]{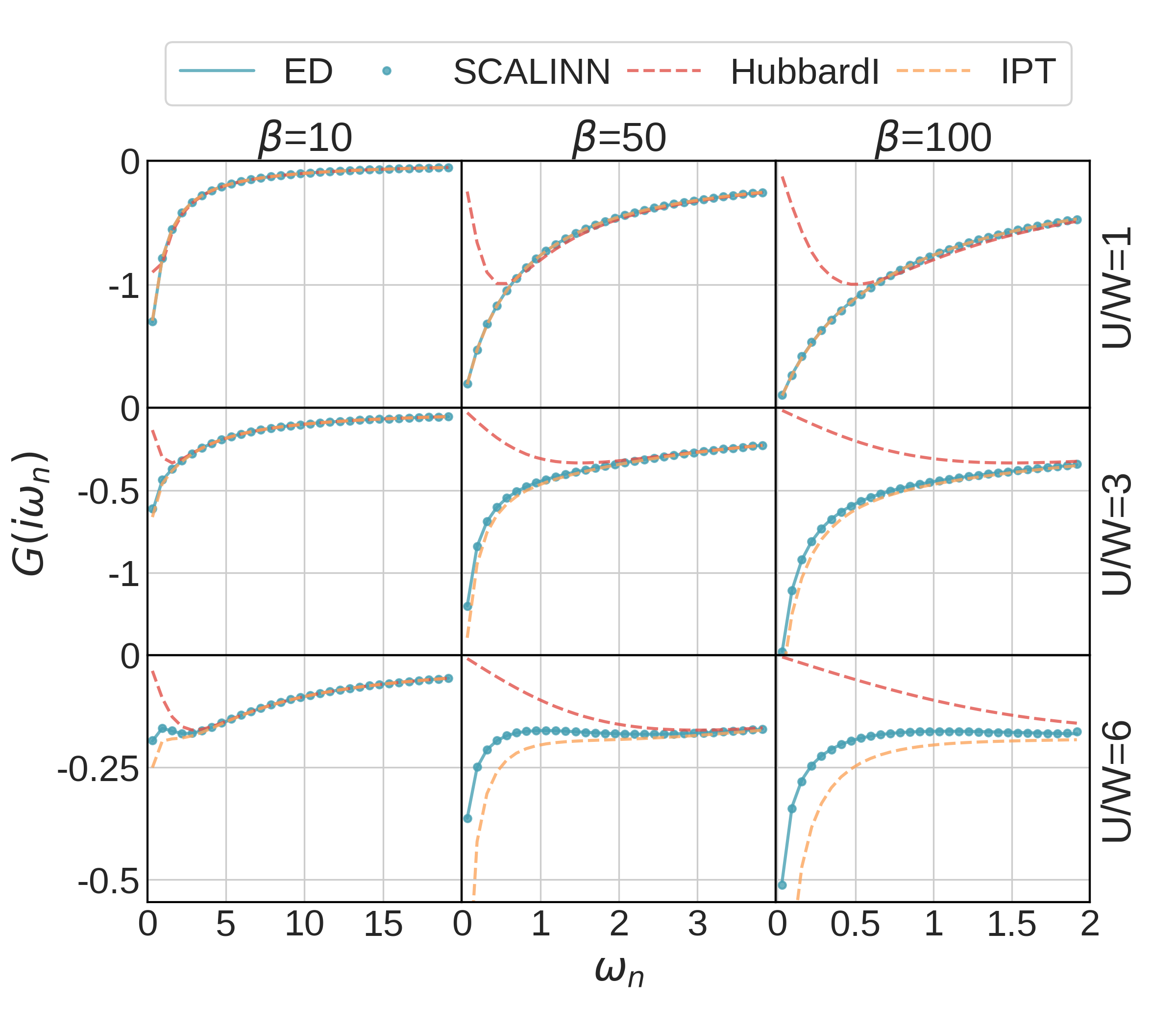}
    \caption{\textbf{Representative SCALINN predictions for a
test-set SIAM with the HI+IPT hybrid scheme.} The Hubbard-I solution $G^{\text{HI}}(i\omega_n)$ (dashed red), the IPT solver solution $G^{\text{IPT}}(i\omega_n)$ (dashed yellow), and the ED target $G^{\text{ED}}(i\omega_n)$ (solid blue lines) are plotted alongside the SCALINN predictions from these input Green's functions (blue dots) at $\beta=10,50,100$ and at $U/W=1,3,6$.}
    \label{fig:hybrid_giw}
\end{figure}

In addition to the truncated bath approach, as discussed in ~\ref{subsec:modes}, we also consider a Hubbard-I+IPT hybrid scheme. In this hybrid method, models were trained from a combination of Hubbard-I and IPT approximate solutions to the target SIAM, and used as inputs to the transformer model to predict target ED-quality outputs. Once again, three separate inverse temperature models were trained to reach errors of $J(\hat{G}(i\omega_n),G^{\text{ED}}(i\omega_n))<10^{-5}$ across the entire training set.

In \cref{fig:hybrid_giw}, the various Matsubara Green's functions are presented, including the Hubbard-I and IPT inputs, the ED ground truth and SCALINN prediction. As expected, the Hubbard-I inputs are all insulating solutions, as observed with $G^{\text{HI}}(i\omega_n)$ tending towards zero as $i\omega_n\rightarrow0$, consistent with its description of the atomic solution. This is increasingly erroneous for lower temperatures and interaction strengths, where delocalized solutions should be found. In contrast, the IPT input favors delocalized descriptions, where all solutions reach a maximum absolute value as $i\omega_n\rightarrow0$, which is in error particularly for higher interactions in the non-perturbative $U/W$ limit. In contrast to these computationally cheap input models, the SCALINN predictions match the ED results to remarkably high accuracy from these inputs. 

\begin{figure}[h!]
    \centering
    \includegraphics[width=0.9\linewidth]{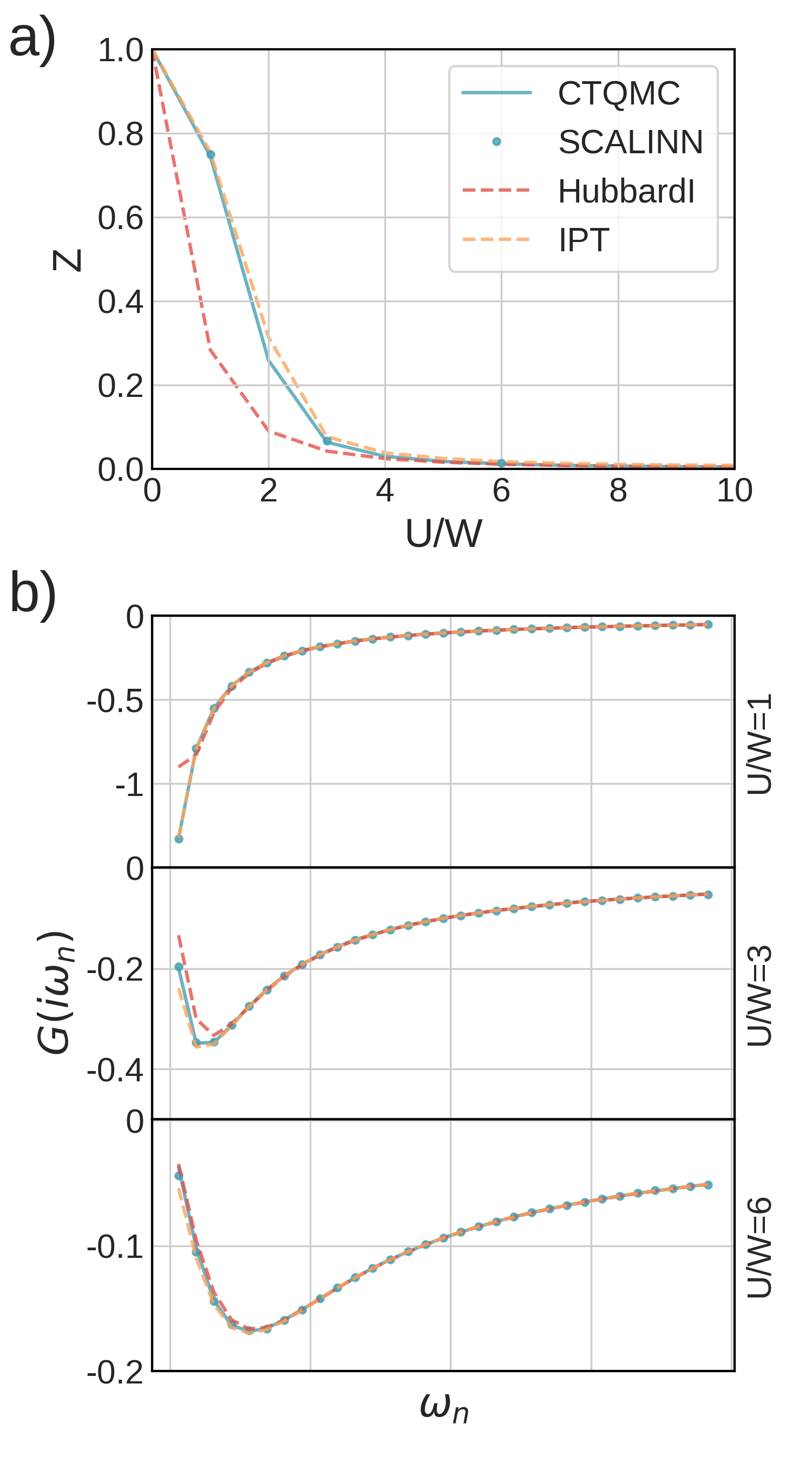}
    \caption{\textbf{Converged self-consistent DMFT results on the Bethe Hubbard lattice for different solvers: IPT, Hubbard-I, SCALINN and CTQMC}. a) Quasi-particle weight b) Imaginary part of Matsubara Green's functions. DMFT is converged at $\beta=10$ for the continuous hybridization of Eq.~\ref{eq:semi-circular} without bath discretization error, with the IPT and Hubbard-I solutions at each iteration used as input for the SCALINN solver approach.}
    \label{fig:hybrid_dmft}
\end{figure}

Finally, we consider the utility of the scheme as a solver within a fully self-consistent DMFT calculation. In this, the IPT and Hubbard-I approximations can be found in the absence of bath discretization error of the hybridization. Therefore, while the training of the SCALINN model is performed in the presence of the finite bath approximation to the hybridization, its use within a DMFT scheme can aim to eliminate both the bath discretization error, as well as approximations to the correlated effects of approximate solvers. In order to benchmark the accuracy of the approach, we therefore turn to comparison with a CTQMC solver which can obtain correlated Green's functions in the limit of a continuous hybridization via Monte Carlo sampling, as long as the temperature is not too low such that the fermion sign problem manifests.

In \cref{fig:hybrid_dmft}, we consider DMFT on the continuous hybridization of the Bethe lattice (Eq.~\ref{eq:semi-circular}) at $\beta=10$ compared to CTQMC, for both the final self-consistent Matsubara Greens function, and the quasiparticle weight, $Z$. This quasiparticle weight can be computed as

\begin{equation}
Z=\left.\frac{\operatorname{Im}\left\{\Sigma\left(i \omega_n\right)\right\}}{\omega_n}\right|_{\omega_n \rightarrow 0}\,,
\end{equation}
from the converged DMFT self-energy. Values close to unity indicate a metallic solution, with lower values describing the increased effective mass towards a Mott insulating solution. This is observed from the results, with DMFT+SCALINN agreeing almost perfectly with the DMFT+CTQMC renormalization factor, with the DMFT+IPT biasing towards the metallic phase, and DMFT+Hubbard-I the atomic Mott phase.

Moreover, the self-consistent DMFT Matsubara Green's functions with these various solvers are plotted in \cref{fig:hybrid_dmft}.b, where as interaction strength increases, the solutions become increasingly insulating, as can be observed from the $i\omega_n\rightarrow0$ trend of $G(i\omega_n)$, with once again the discrepancy between SCALINN and CTQMC solvers is indistinguishable on the scale of the plot.
It should be noted that these insulating solutions differ from the single-shot SIAM solutions of \cref{fig:hybrid_giw}, due to the self-consistent update of the continuous hybridization in the full DMFT scheme.

\section{Conclusions}
With the insights gained from drawing comparisons to natural language processing problems, we developed a novel and promising approach to predict Green's function sequences in the Matsubara domain via modifications to a Transformer model.  These predicted sequences exhibit levels of accuracy that were previously restricted to comparatively high computational costs of exact diagonalization. We considered approaches to both predict these Green's functions of general SIAM models from inputs based on results from computationally accessible lower levels of theory, as well as an approach to mitigate the bath discretization error in describing SIAM's with a continuous hybridization spectrum. Finally, we combined these developments in a fully self-consistent DMFT scheme to solve the Bethe lattice Hubbard model with results indistinguishable from exact CTQMC benchmarks. 

However, while the approach showcases much potential, there exist remaining challenges to overcome. Firstly, the approach was restricted to a relatively narrow class of SIAM models from a training dataset of a maximum of 7 fermionic levels. This set of training data can still be improved upon to reach competitive levels as a DMFT solver.As such, the extension to multi-impurity, matrix-valued Green's functions and a wider set of representative hybridizations is required. The advantage gained from the use of this model is that once a model is trained from this set of larger number of fermionic levels, wider hybridization representation extended dataset, the run-time GPU computational cost does not increase accordingly. Note however that one of the key take-aways of our work is that sampling the entire space of AIM (those represented by 7-bath sites and all those that aren't) is not required to achieve excellent accuracy for the machine learning approach. Essentially our work demonstrates that training the network on discretized AIMs provides a solid network that can apply to any AIMs. One remarkable demonstration is that comparison to the exact (within statistical error bars) CTQMC results for DMFT. Extending the database with AIM with larger bath sites has been shown in our work to only provide minor improvements, see for instance the analysis in \cref{eq:delta2} in the appendix where we review the dependence on number of bath sites. Adaptations of the model to {\em enforce} desirable features such as causality of the output Green's functions or symmetries would be beneficial, and help avoid convergence issues in the self-consistent DMFT loops which manifested at times for low temperatures or quantum phase transitions. Finally, alternative methods to provide training data of exact Green's functions would allow for an extension to overcome the bath discretization which is manifest in the training of the model. Nevertheless, our findings highlight the power and adaptability of the Transformer model within the field of correlated materials and its potential for pushing the frontiers of computational problem-solving in this domain.

\section*{Acknowledgment}
G.H.B. gratefully acknowledges support from the Air Force Office of Scientific Research under award number FA8655-22-1-7011.
We are also grateful to the King's Computational Research, Engineering and Technology Environment (CREATE) and UK Materials and Molecular Modelling Hub for computational resources, which is partially funded by EPSRC (EP/T022213/1, EP/W032260/1 and EP/P020194/1).

\section*{COMPETING INTERESTS}
The authors declare no competing financial or non-financial interests.

\section*{Contributions}
C.W., G.H.B, W.G., and Z.Z. conceived the project. C.W. and Z.Z build the database. W.G., H.L and Z.Z developed the Machine Learning model and trained the Transformer model. H.L and Z.Z. performed data analysis. All wrote the manuscript.

\section*{Corresponding author}
Correspondence to Weifeng Ge and Cedric Weber

\section*{Data availability}
An exemplary dataset can be found at \url{https://dx.doi.org/10.6084/m9.figshare.23144474}.

\section*{Code availability}
The code for SCALINN is available at \url{https://github.com/zelong-zhao/SCALINN}.

\bibliographystyle{apsrev4-2}
% \bibliographystyle{sn-nature}
% \printbibliography
% \bibliography{bib_for_ml}

%apsrev4-2.bst 2019-01-14 (MD) hand-edited version of apsrev4-1.bst
%Control: key (0)
%Control: author (72) initials jnrlst
%Control: editor formatted (1) identically to author
%Control: production of article title (-1) disabled
%Control: page (0) single
%Control: year (1) truncated
%Control: production of eprint (0) enabled
%

\clearpage

\clearpage

\appendix 

%%%%%%%%%% Merge with supplemental materials %%%%%%%%%%
\widetext
\begin{center}
\textbf{\large Supplemental Materials: A language-inspired machine learning approach for solving strongly correlated problems with dynamical mean-field theory}
\end{center}

\setcounter{equation}{0}
\setcounter{figure}{0}
\setcounter{table}{0}
\setcounter{page}{1}
\setcounter{section}{0}
\makeatletter
\renewcommand{\theequation}{A\arabic{equation}}
\renewcommand{\thefigure}{A\arabic{figure}}
\renewcommand{\thetable}{A\arabic{table}}

\section{Hyperparameter optimisation}

\subsection{Transformer Hyperparameter Tuning}

\noindent \textbf{Task:} \\

\begin{sidewaysfigure}
    \centering
    \includegraphics[width=1.\textheight,height=0.5\textwidth]{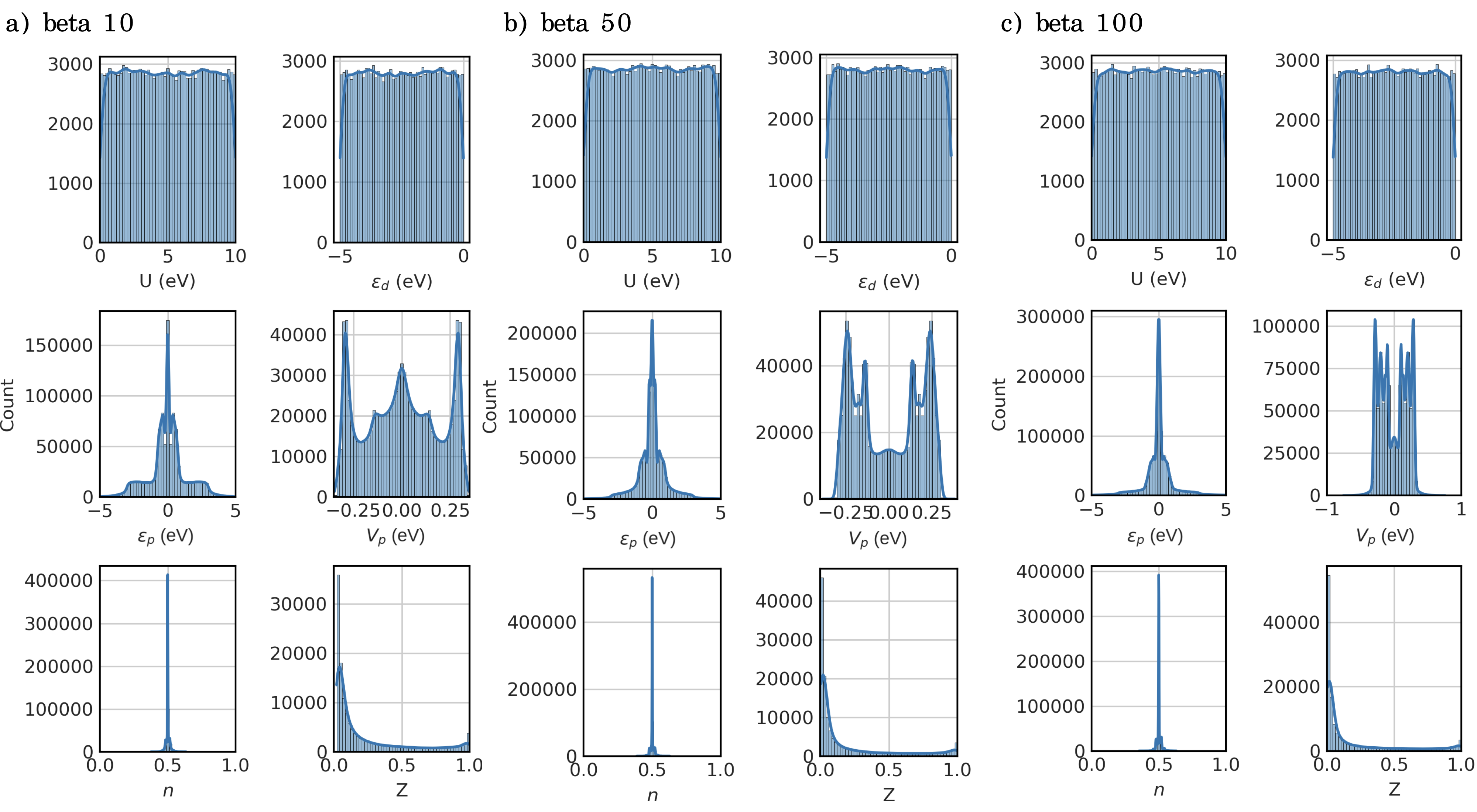}
    \caption{Database of parameters are sampled from a maximum radius of 1~eV from semicircular Green's function, Y axis is the count of the samples. The symbols $U$ and $\epsilon_d$ refer to the impurity Hubbard $U$ term and on-site energy, respectively. On the other hand, $\epsilon_p$ and $V_p$ denote the on-site energy and hopping amplitude, respectively. a) b) c) are sampled from different temperatures.}
    \label{fig:semi_cir_cular_training-dataset}
\end{sidewaysfigure}

The purpose of the Transformer model is to recover the imaginary component of the Green's function $\mathrm{Im}[G(i\omega_n)] \in \mathbb{R}^{N_\omega}, n \in [0,N_\omega-1], N_\omega=31$, as calculated with a 7-bath exact diagonalization (ED) solver, by learning the information from truncated 5-bath ED solutions. This is carried out though fine-tuning the Transformer model parameters as listed below. Unless stated otherwise, the models were trained with the full database as shown in \cref{fig:semi_cir_cular_training-dataset}. The samples of the database were split such that 80\% of the data was allocated for training, and 20\% for validation and the many-body problem were solved at default values of inverse temperature $\beta$ of 10, 50, and 100. \\

\noindent \textbf{Model Default Settings:} \\

Following the details of the main text, the positional encoding of the Transformer network is replaced by the stacking of the Matsubara frequency. The application of the Gaussian Error Linear Unit (GELU) activation function has been incorporated into the model. In the following sections, we also analyze other extensions of this idea of positional encoding via Matsubara frequency. Moreover, in relation to the masks of the decoder, the look-forward (LF) and look-backward (LB) masks are limited to five time steps before and after a particular sequence entry respectively. We utilized a batch size of 8 and a learning rate of $10^{-7}$, with the Adam optimizer employed for training. Each task was performed three times, from which the metrics corresponding to the best performing model are reported. 

\begin{table}[h!]
\caption{Transformer model performance test. Unlisted value are identical to those of base value. N$_{\mathrm{model}}$ is the number of encoder and decoder blocks, n$_{Encoder/Decoder}$,within the Transformer. d$_{\mathrm{model}}$ denotes the dimensions of the multilayer perceptron blocks (MLP) and self-attention layers in the model. h specifies the number of heads. d$_{k/v}$ is the dimensions of the heads in the multi-head self-attention block. The term d$_{\mathrm{ff}}$ refers to the dimensions of the feedforward layers in the Transformer. In the mask column, LF and LB are abbreviations for look-forward and look-backward, respectively. In the input transformation column, N denotes normalization of data sets to between 0 and 1, and Z represents standardization of data.}
\label{tab:beta100_truncated5}
\begin{tabular}{c|cccccccccc|ccc}
\hline
 & N$_{\mathrm{model}}$ & d$_{\mathrm{model}}$ & d$_{\mathrm{ff}}$ & h & d$_{\mathrm{k/v}}$ & mask & PE & \begin{tabular}[c]{@{}c@{}}Inputs\\ Transform\end{tabular} & dropout & Train-Step & \begin{tabular}[c]{@{}c@{}}MSE$_{\beta 100}$\\ \,\end{tabular} & \begin{tabular}[c]{@{}c@{}}MSE$_{\beta 50}$ \\ \,\end{tabular} & \begin{tabular}[c]{@{}c@{}}MSE$_{\beta 10}$ \\ (Val) ($10^{-6}$)\end{tabular} \\ \hline
Base & 4 & 256 & 1024 & 32 & 8 & LF5 LB5 & \begin{tabular}[c]{@{}c@{}}Matsu\\ Freq($\oplus$)\end{tabular} & N/A & 0. & 350K & 22.21 & 7.57 & \textbf{4.74} \\ \hline
(a.1) &  &  &  & 4 & 64 &  &  &  &  &  & 37.31 & 8.13 & 5.03 \\
(a.2) &  & 128 & 512 & 16 &  &  &  &  &  &  & 31.18 & 14.37 & 6.02 \\
(a.2) & 2 & 64 & 128 & 8 &  &  &  &  &  &  & 46.30 & 21.39 & 12.99 \\ \hline
 &  &  &  &  &  &  & Learned(add) &  &  &  & 2,095.49 & 660.43 & 756.03 \\
(b) &  &  &  &  &  &  & Learned($\oplus$) &  &  &  & 57,474.60 & 36,406.33 & 13,774.82 \\
 &  &  &  &  &  &  & Cosine(add) &  &  &  & 5,597.37 & 7,055.71 & 15,538.61 \\ \hline
 &  &  &  &  &  & LF1 LB1 &  &  &  &  & 34.33 & 8.10 & 5.25 \\
(c) &  &  &  &  &  & LF3 LB3 &  &  &  &  & 28.95 & 8.62 & 5.49 \\
 &  &  &  &  &  & LF10 LB10 &  &  &  &  & 27.62 & 8.36 & 6.06 \\ \hline
 &  &  &  &  &  &  &  &  & 0.01 &  & 53.29 & N/A & N/A \\
(d) &  &  &  &  &  &  &  &  & 0.05 &  & 214.91 & N/A & N/A \\
 &  &  &  &  &  &  &  &  & 0.1 &  & 581.48 & N/A & N/A \\ \hline
(e) &  &  &  &  &  &  &  & N &  &  & 6.30 & 2.70 & 3.98 \\
 &  &  &  &  &  &  &  & Z &  &  & 300.61 & 189.57 & 242.42 \\ \hline
(a.1) & 6 & 512 & 2048 & 64 &  &  &  &  &  &  & \textbf{19.77} & 7.93 & 5.21 \\
Big & 6 & 512 & 2048 & 8 & 64 &  &  &  &  &  & 21.63 & \textbf{7.37} & 4.78 \\ \hline
\end{tabular}
\end{table}

\noindent \textbf{Tab. \ref{tab:beta100_truncated5} (a): Model Size Test }

\noindent We conducted a model size test experiment using the model default settings and task mentioned above. In addition to the base model, alternative settings of the Transformer size were tested. Each test was performed three times, and the results with the lowest loss are presented. 

\noindent Across different temperatures, the shallower and narrower Transformer model (a.2) converges more slowly compared to models with larger dimensions. Models with additional Transformer encoder and decoder layers, or increased dimensions, tend to deliver superior performance. The dimension of the model's attention heads (a.1) has a more complex influence on model performance: For the base model, larger head dimensions degrade the model's performance. However, for larger models, increased head dimensions reduce performance at low temperatures but improve performance at high temperatures.

\vspace{0.5cm}

\noindent \textbf{Tab. \ref{tab:beta100_truncated5} (b): Positional Encoding Methods Test}

\noindent We conducted a positional encoding experiment using the model default settings and task mentioned above. In addition to the base model, alternative positional encoding (PE) methods were explored. Four distinct positional encoding approaches were evaluated: learned PE ($\oplus$), learned PE (add) and stacking of the Matsubara frequencies ($\oplus$). These methods involve either trained or untrained parameters being stacked or added to the truncated inputs $\mathrm{Im}[G(i\omega_n)^{trunc}_l]$. In contrast, the original cosine PE is added to the encoded $\mathrm{Im}[G(i\omega_n)^{trunc}_l]$. Each test was performed three times, and the results with lowest loss are presented. Overall, the stacking of Matsubara frequency was found to be the optimal method across all values of beta.

\noindent 

\vspace{0.5cm}

\noindent \textbf{Tab. \ref{tab:beta100_truncated5} (c): Mask Shape Test}

\noindent We conducted a mask shape test experiment using the model default settings and task mentioned above. In addition to the base model, alternative settings of mask were tested. Four sets of values are investigated: 1, 3, 5, and 10 corresponding to the quantity of unmasked (zero) diagonal matrix lines, with the rest of the matrix entries as negative infinity. Equal steps are taken for both look forward and look backward. In the Transformer's decoder-encoder self-attention layer mask, both look-forward and look-backward are employed. However, only look-backward is applied in the decoder input self-attention layer. Each initialized model is then trained using these distinct mask shapes and the results with lowest loss are presented.

\noindent Across different temperatures, neither a small (LF1LB1) nor a large (LF10LB10) scope of unmasked inputs achieved optimal performance. At a beta of 100, increasing the unmasked area from LF1LB1 to LF5LB5 improved model performance, but with LF10LB10, the performance declined. At higher temperatures, increasing the unmasked area initially decreased model performance. However, a performance maximum appears around a masked size of (LF5LB5) for all temperatures. This suggests that overall, a moderate size mask of LF5LB5 should be adopted for the Transformer model.

\vspace{0.5cm}

\noindent \textbf{Tab. \ref{tab:beta100_truncated5} (d): Dropout Test}

\noindent We conducted a dropout test using the model default settings and task mentioned above. In addition to the base model,  three dropout values, 0.01, 0.05 and 0.1 are tested in input layer and the Transformer encoder and decoder layers. In this test, only the case of $\beta$=100 is considered, as lower temperatures tend to be more challenging to fit for such tasks. Three separate models are randomly initialized to minimize noise arising from the initial configuration. Each initialized model is subsequently trained using varying dropout rate. Overall, across all dropout values tested in various positions within the model, excluding dropout leads to a significant performance improvement compared to including dropout.

\vspace{0.5cm}

\noindent \textbf{Tab. \ref{tab:beta100_truncated5} (e): Green's function transform}

\noindent We conducted a inputs transform tests using the model default settings and task mentioned above. In addition to the base model, two choices of inputs transform were tested. There are several choices of transformation factors for the target since it is a 7-bath SIAM solved by ED, while the inputs are truncated solutions with 5-\st{bath sites}\textcolor{red}{bath orbitals}. We use transformation factors derived from the target, which is the ED of the 7-bath system. It should be noted that the input Green's function is not transformed as the name suggests. However, the output of the Transformer model is either normalized or standardized. Each test was performed three times, and the results with lowest loss are presented. 

\noindent In summary, across different temperatures, the performance of the model is negatively impacted by standardizing the Green's function. As the temperature increases, models using normalized Green's functions show better performance compared to those using the raw physical values of the Green's functions.

\clearpage

\subsubsection*{Batch LR}

\par
\begin{figure}[h!]
    \centering
    \includegraphics[width=0.9\linewidth]{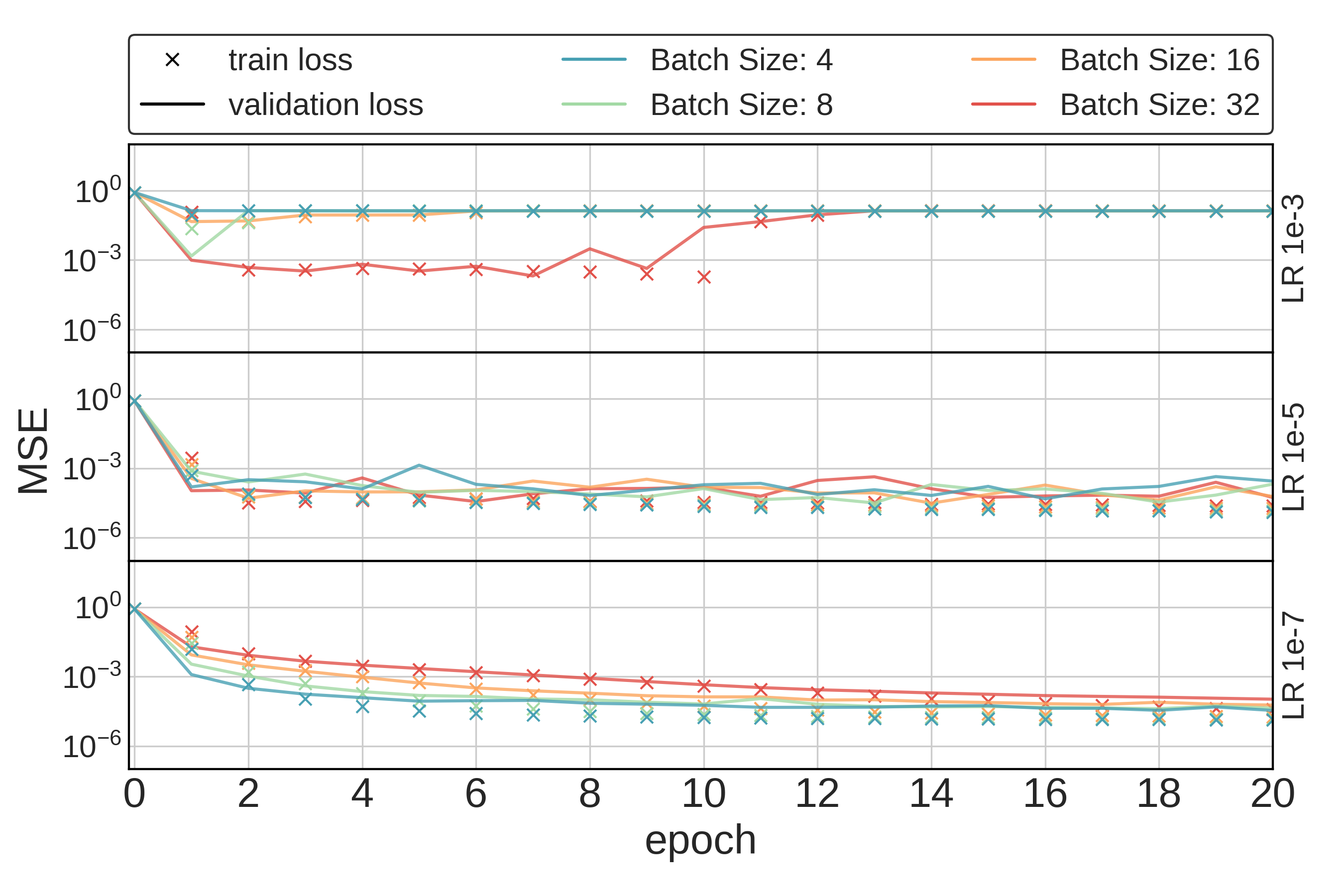}
    \caption{Performance of the Transformer network as a function of batch size and learning rate at an inverse temperature $\beta$ value of 100. Dashed lines represent training loss, while solid lines represent validation loss.}
    \label{fig:batch_lr_loss_MSE}
\end{figure}

 We conducted a batch size and learning rate test using the base model and task mentioned above. In addition to the \textbf{model default settings}, we tested three learning rates, 1e-3, 1e-5, and 1e-7, along with four batch sizes, 4, 8, 16, and 32. In this test, only the case of $\beta$=100 is considered, as lower temperatures tend to be more challenging to fit for such tasks. Three separate models are randomly initialized to minimize noise arising from the initial configuration. Each initialized model is subsequently trained using varying batch sizes and learning rates, and the average results are illustrated in Figure \ref{fig:batch_lr_loss_MSE}.

With a high learning rate of 1e-3, the loss function converges to a relatively high, yet similar value across all batch sizes. When the learning rate is decreased to 1e-5, a lower loss is achieved across all batch sizes, however, the impact of varying batch size on performance is not distinctly clear. At a learning rate of 1e-7, the model can be trained effectively. However, it is noteworthy that training the model becomes more difficult with larger batch sizes, although they eventually converge to similar values. While larger batch sizes expedite the training process on a per iteration basis, they cause a slower decrease in the loss function. Consequently, a mid-range batch size is recommended for achieving an optimal balance between training efficiency and model performance.

\clearpage

\subsection{Database Size}

\par
\begin{figure}[h!]
    \centering
    \includegraphics[width=1.\linewidth]{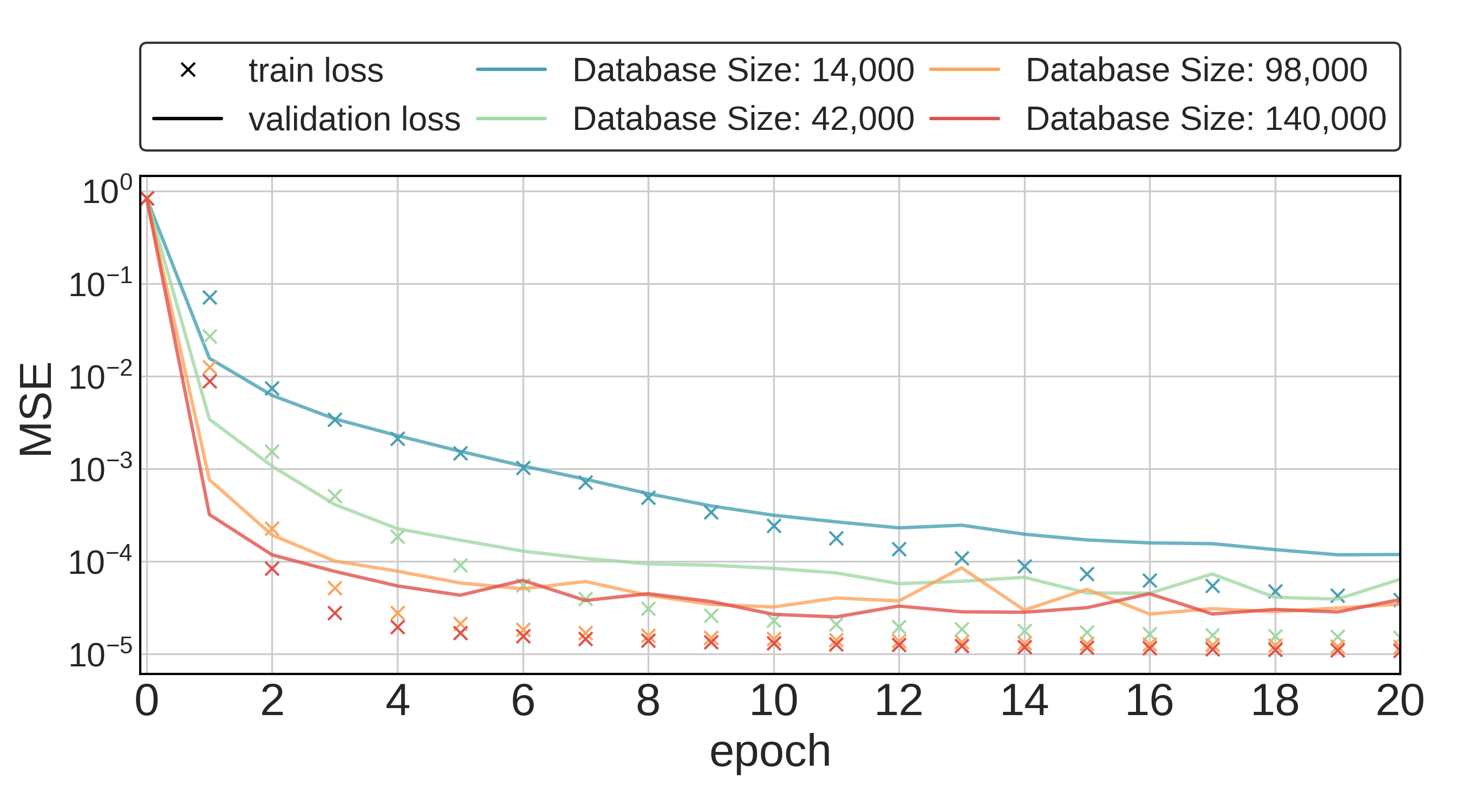}
    \caption{Model performance as a function of database size at an inverse temperature value of 100. Database size indicates the number of sum of training and validation samples of the outputs, of which 80 percent was used for training and 20 percent for validation. The dashed lines represent the training loss, while the solid lines represent the validation loss.}
    \label{fig:dtabase_size}
\end{figure}

 We conducted a database size test using the model default settings and base model. In addition to the \textbf{task} mentioned above, three size of training sets were tested. In this test, only the case of $\beta$=100 is considered, as lower temperatures tend to be more challenging to fit for such tasks. Three separate models are randomly initialized to minimize noise arising from the initial configuration. Each initialized model is subsequently trained using varying settings, and the average results are illustrated in Figure \ref{fig:dtabase_size}.

\par At a beta of 100, increasing the number of training samples does not appear to lower the validation loss for this particular task. This could potentially be due to the saturation of unique features in the larger dataset. Hence, in a larger database, similar samples would be trained more frequently within a single epoch compared to a smaller database, which doesn't necessarily contribute to a further decrease in validation loss. 

\clearpage

\subsection{Multimodel}

\par In this section, we explore various methods to incorporate Anderson impurity model (AIM) parameters into the Transformer model, using the tests and model default settings previously mentioned. We categorize the AIM parameters into hybridization parameters and impurity parameters. For the impurity parameters, a straightforward approach of using a MLP is adopted. However, for the hybridization parameters, we implement a more complex configuration: two row-wise MLPs connected by mean pooling. This approach is designed to maintain the permutation symmetry properties associated with the \st{bath sites}\textcolor{red}{bath orbitals}.

\par In addition to model size, there are several options for merging encoded parameters into the Transformer model. These options including how to encode these parameters and where to merge them. Regarding location, we tested two scenarios: one where the encoded parameters are stacked with the positional encoder, and another where they are stacked to the output of the Transformer encoder. Furthermore, in the task of recovering a 7-bath SIAM Hamiltonian using a 5-bath solution, three Hamiltonians need to be encoded. Consequently, we also evaluated various pooling methods to enhance model performance.

\begin{table}[h!]
\caption{Parameter encoder model performance test. Unlisted value are identical to those of base value. The base model corresponds to the model in (Tab. \ref{tab:beta100_truncated5}). `N base' stands for Normalised Base, which is identical to the model described in (Tab. \ref{tab:beta100_truncated5} (d)). In this test, the Transformer model size is also set to the base size. "ParamMergePos" is an abbreviation for the encoded parameters merging position in the Transformer. In this column, "PE" represents the position where positional encoding is included in the inputs. "KV" refers to the self-attention layer in the Transformer decoder (base model), where KV is sourced from the Transformer encoder. The "Pool method" refers to the technique employed to facilitate interaction between processed hybridization inputs and target hybridization before they are directed into the Transformer decoder.}
\label{tab:beta100_truncated5_enc_param}
\begin{tabular}{c|cccccccc|ccc}
\hline
 & N$_{\mathrm{Hyb}}$ & N$_{\mathrm{Imp}}$ & d$_{\mathrm{model}}$ & d$_{\mathrm{out}}$ & \begin{tabular}[c]{@{}c@{}}Inputs\\ Trasnform\end{tabular} & \begin{tabular}[c]{@{}c@{}}Param\\ Merge\\ Pos\end{tabular} & \begin{tabular}[c]{@{}c@{}}Pool\\ method\end{tabular} & Train-Step & \begin{tabular}[c]{@{}c@{}}MSE$_{\beta 100}$\\ \,\end{tabular} & \begin{tabular}[c]{@{}c@{}}MSE$_{\beta 50}$\\ \,\end{tabular} & \begin{tabular}[c]{@{}c@{}}MSE$_{\beta 10}$\\ (Val) ($10^{-6}$)\end{tabular} \\ \hline
Base & N/A & N/A & 256 & N/A & N/A & N/A & N/A & 350K & 22.1 & 7.57 & 4.74 \\ \hline
 & 1 & 1 &  & 32 &  & PE &  &  & 5,949.96 & 941.17 & 63.57 \\
(a) & 2 & 2 &  & 64 &  & PE &  &  & 5,034.78 & 2386.36 & 581.99 \\
 & 4 & 4 &  & 128 &  & PE &  &  & 20,781.07 & 1670.32 & 182.43 \\ \hline
 & 1 & 1 &  & 32 & N & KV & Mean &  & \textbf{2.99} & 2.80 & 2.98 \\
(b) & 2 & 2 &  & 64 & N & KV & Mean &  & 7.64 & 2.93 & 3.14 \\
 & 4 & 4 &  & 128 & N & KV & Mean &  & 8.47 & 3.04 & 4.43 \\ \hline
 & 1 &  &  & 32 & N & KV & Mean &  & 8.35 & \textbf{2.63} & 3.99 \\
(c) & 2 &  &  & 64 & N & KV & Mean &  & 7.91 & 3.12 & 4.08 \\
 & 4 &  &  & 128 & N & KV & Mean &  & 7.82 & 3.18 & 4.28 \\ \hline
 &  & 1 &  & 32 & N & KV &  &  & 6.56 & 2.71 & \textbf{2.94} \\
(d) &  & 2 &  & 64 & N & KV &  &  & 8.37 & 3.0 & 3.69 \\
 &  & 4 &  & 128 & N & KV &  &  & 7.11 & 2.91 & 4.45 \\ \hline
 & 2 & 2 &  & 64 & N & KV & $\oplus$ &  & 7.47 & 3.55 & 3.84 \\
(e) & 2 & 2 &  & 64 & N & KV & Minus &  & 6.58 & 3.44 & 3.78 \\
 & 2 & 2 &  & 64 & N & KV & add &  & 9.10 & 3.32 & 3.90 \\ \hline
N base &  &  &  &  & N &  &  &  & 6.30 & 2.70 & 3.98 \\ \hline
\end{tabular}
\end{table}

\noindent \textbf{Tab. \ref{tab:beta100_truncated5_enc_param} (a): Integrating Encoded AIM Parameters with Positional Encoder}

\noindent We conducted a model size test for the parameter encoder using the model default settings and task mentioned above. In addition to the base model, encoded parameters were stacked alongside positional encoding. Three model sizes were tested. Meanwhile, mean pooling was utilized to handle multiple hybridization inputs. Each test was performed
three times, and the results with lowest loss are presented. Overall, stacking encoded Hamiltonian parameters with Positional Encoding (PE) to the Transformer led to a reduction in the Transformer's performance across all tested temperature.

\vspace{0.5cm}

\noindent \textbf{Tab. \ref{tab:beta100_truncated5_enc_param} (b): Integrating Encoded AIM Parameters in Decoder's Self-Attention Layer}

\noindent We conducted a model size test for the parameter encoder using the model default settings and task mentioned above. In addition to the base model, encoded parameters were stacked the Transformer decoder. Three model sizes were tested in each case. Meanwhile, mean pooling was utilized to handle multiple inputs of hybridization. Each test was performed three times, and the results with the lowest loss are presented. Overall, one layer of the parameter encoder achieved the optimal results.

\vspace{0.5cm}

\noindent \textbf{Tab. \ref{tab:beta100_truncated5_enc_param} (c,d): Ablation Study of Multi-Model in Decoder's Self-Attention Layer}

\noindent In addition to (Tab. \ref{tab:beta100_truncated5_enc_param} (b)), we investigated how separately encoding the impurity parameters and hybridization parameters impacts the performance of the Transformer across all temperature ranges.

\noindent Larger model sizes for the impurity parameter encoder resulted in poorer performance. The situation for the hybridization parameter encoder was more complex. At lower temperatures, larger models tended to have better performance, while at higher temperatures, larger models performed worse. Interestingly, when only the impurity or hybridization parameter encoder was used, the model performance was sub-optimal (worse than 'N base'). However, when both were used, the model performance improved at beta of 100.

\vspace{0.5cm}

\noindent \textbf{Tab. \ref{tab:beta100_truncated5_enc_param} (e): Multi-hybridization Pooling Methods Test}

\noindent We conducted pooling method test for the parameter encoder using the model default settings and task mentioned above. In addition to the previous  model, different pooling methods were tested to handle multiple inputs of hybridization. Each test was performed three times, and the results with the lowest loss are presented. Overall, subtracting different hybridizations was found to be the optimal method for low temperatures, and meaning pooling have better performance at high temperatures. 

\subsection{Inputs test}
In this section we explore changes in model performance when taking different forms of input at different temperatures. Training settings followed the default setting mentioned above but the task inputs were changed to different schemes, where different truncated bath size and different approximation solutions were tested. 

\begin{table}[h!]
\caption{N$_{\mathrm{Hyb}/ \mathrm{Imp}}$ refers to the number of layers in the parameter encoder, where the dimensions of the parameter encoder equal the number of layers multiplied by 32. For models implementing the parameter encoder, the outputs of the parameter encoder are stacked with the Transformer encoder output before being fed into the Transformer decoder's encoder layer, as shown in Table \ref{tab:beta100_truncated5_enc_param} (b). \st{G$_{\mathrm{in}}$}\textcolor{red}{$G(i\omega_n^\mathrm{input})$} represents the source of the inputs directed into the Transformer model and N$_{\mathrm{G}}$ stands for number of source inputs. }
\label{tab:inputs-test}
\begin{tabular}{c|cccccc|ccc}
\hline
 & N$_{\mathrm{Hyb}/ \mathrm{Imp}}$ & d$_{\mathrm{model}}$ & \textcolor{red}{$G(i\omega_n^\mathrm{input})$} & N$_G$ & \begin{tabular}[c]{@{}c@{}}Inputs\\ Transform\end{tabular} & Train-Step & \begin{tabular}[c]{@{}c@{}}MSE$_{\beta 100}$\\ \,\end{tabular} & \begin{tabular}[c]{@{}c@{}}MSE$_{\beta 50}$\\ \,\end{tabular} & \begin{tabular}[c]{@{}c@{}}MSE$_{\beta 10}$\\ (Val) ($10^{-6}$)\end{tabular} \\ \hline
Base & N/A & 256 & \textcolor{red}{truncation of 2 orbitals} & 3 & N/A & 350K & 22.1 & 7.57 & 4.74 \\ \hline
 & 2 &  &  &  & N &  & \textbf{7.64} & 2.94 & \textbf{3.14} \\
(a) & 2 &  & \textcolor{red}{truncation of 3 orbitals} &  & N &  & 9.98 & \textbf{2.81} & 3.32 \\
 & 2 &  & \textcolor{red}{truncation of 4 orbitals} &  & N &  & 10.97 & 3.10 & 3.24 \\ \hline
 &  &  & IPT \& HI & 2 &  &  & 18.32 & 6.29 & 4.73 \\ \hline
 & 2 &  & HI & 1 & N &  & 7.75 & 2.02 & \textbf{2.28} \\
(b) & 2 &  & IPT & 1 & N &  & \textbf{4.98} & \textbf{1.82} & 2.42 \\
 & 2 &  & IPT \& HI & 2 & N &  & 5.62 & \textbf{1.82} & 2.37 \\ \hline
\label{tab:bathsize}
\end{tabular}
\end{table}

\noindent \textbf{Tab. \ref{tab:inputs-test} (a): Truncated Bath Size Test}

\noindent We conducted tests with varying truncated bath sizes, using the model default settings mentioned above. This refers to the number of baths extracted from the target bath. Both AIM parameters and the imaginary part of the Green's function were used as inputs. Generally, a smaller truncated bath size tends to contain less information about the entire system. Each test was performed three times, and the results with the lowest loss are presented. As expected, the Transformer model tends to perform better with larger bath sizes.

\vspace{0.5cm}

\noindent \textbf{Tab. \ref{tab:inputs-test} (b): Transformer with Approximation Solver Test}

\noindent We conducted inputs test with different inputs using model default settings mentioned above. We take both AIM parameters and imaginary part of Green's function as inputs. Each test was performed three times, and the results with the lowest loss are presented.

\noindent At low temperatures, the model that used IPT as inputs outperformed the model that used HI inputs. Moreover, the model that incorporated both IPT and HI (hybrid) solvers demonstrated accuracy comparable to the best-performing model at moderate temperatures. At other temperatures, its performance was intermediate between models that used only one of the hybrid solvers.

\section{Database Sampling}
Here we present the distribution of the datasets used for training and validation. The dataset is collected with random sampling of 1\,eV of semi-circular DOS (Eq. \ref{eq:semi-circular}) at half-filling. Density and quasi-particle weight distribution are found by ED of 7-\st{bath sites}\textcolor{red}{bath orbitals}. $U$ is randomly sampled from 0 to 10\,eV. The impurity level $\epsilon_d$ is kept at $-U/2$. THe on-site bath level, $\epsilon_p$ is sampled between -5 to 5\,eV, and the hopping term $V_p$ changes by temperature. The results of such sampling did not reproduce perfect half-filled solutions for all the SIAM samples, and as temperature decrease more samples deviate from being exactly half-filled.

\begin{figure}[h!]
    \centering
    \includegraphics[width=1.\linewidth]{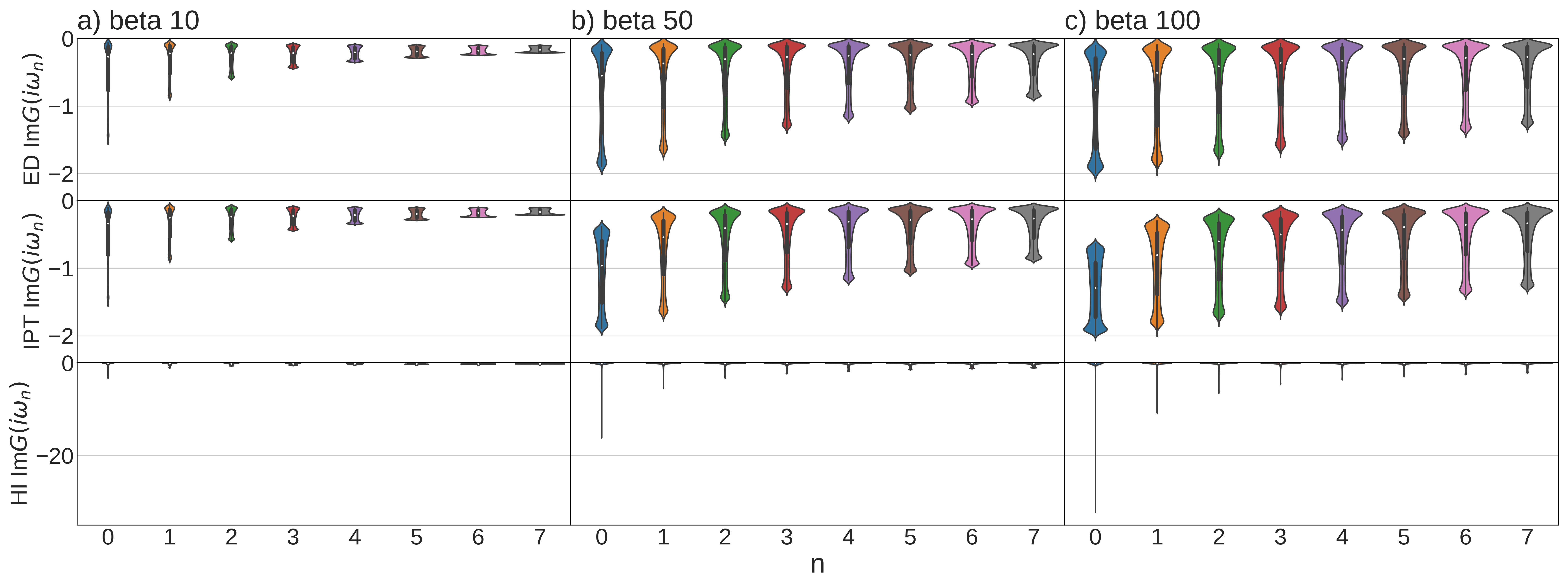}
    \caption{Green's function found by different solver by given SIAM parameters at different temperature. n is the Matsubara frequency index. Here, the distribution of first 8 $G(i\omega_n)$ points are presented due to their significance in determine the low energy properties of the solution. Distributions at different frequencies are shown as violin plots which indicate their density distributions. A white dot is present within each distribution to denote the median value. While the black boxes illustrate the interquartille range(IQR), and lastly, black lines represent $1.5\times$ the interquartile range.}
    \label{fig:my_label}
\end{figure}

\end{document}